\documentclass[cernpreprint, atlasdraft=false, USenglish, texmf, orcidlogo]{atlasdoc}
\usepackage{atlaspackage}
\usepackage{graphicx}
\usepackage{subfig}
\usepackage{siunitx}
\usepackage[block=space]{atlasbiblatex}
\usepackage{comment}

\usepackage{atlasphysics}
\graphicspath{{logos/}{figures/}}

\usepackage{ANA-HDBS-2018-39-PAPER-defs}

\usepackage{pdflscape}
\usepackage{pifont}
\usepackage{tikz}
\usetikzlibrary{positioning}
\usepackage{siunitx}

\AtlasTitle{Search for charged Higgs bosons produced in top-quark decays or in association with top quarks and decaying via $H^{\pm} \to \tau^{\pm}\nu_{\tau}$ in 13~TeV $pp$ collisions with the ATLAS detector}

\AtlasAbstract{%
Charged Higgs bosons produced either in top-quark decays or in association with a top quark,
subsequently decaying via $H^{\pm} \to \tau^{\pm}\nu_{\tau}$, are searched for in 140~$\text{fb}^{-1}$ of
proton--proton collision data at $\sqrt{s}=13$~TeV recorded with the ATLAS detector. Depending
on whether the top quark is produced together with the $H^{\pm}$ decays hadronically or semi-leptonically,
the search targets $\tau$+jets or $\tau$+lepton final states, in both cases with a $\tau$-lepton decaying into a neutrino and hadrons.
No significant excess over the Standard Model background expectation is observed.
For the mass range of $80 \leq m_{H^{\pm}} \leq 3000$~GeV,
upper limits at 95\% confidence level are set on the production cross-section of the charged
Higgs boson times the branching fraction $\mathrm{\cal{B}}(H^{\pm} \to \tau^{\pm}\nu_{\tau})$ in
the range 4.5~pb--0.4~fb. In the mass range 80--160~GeV, assuming the Standard Model cross-section
for $t\bar{t}$ production, this corresponds to upper limits between 0.27\% and 0.02\% on
$\mathrm{\cal{B}}(t\to bH^{\pm}) \times \mathrm{\cal{B}}(H^{\pm} \to \tau^{\pm}\nu_{\tau})$.}

\AtlasRefCode{HDBS-2018-39}

\PreprintIdNumber{CERN-EP-2024-311}

\arXivId{2412.17584}

\HepDataRecord{ins2862529}

\AtlasJournal{Physical Review D}
\AtlasJournalRef{\PRD 111 (2025) 072006}
\AtlasDOI{10.1103/PhysRevD.111.072006}

\hypersetup{pdftitle={ATLAS document},pdfauthor={The ATLAS Collaboration}}

\clearpage
\def\Hp{\ensuremath{H^+}}
\def\Hm{\ensuremath{H^-}}

\def\bbar{\ensuremath{{\overline b}}}
\def\tbar{\ensuremath{{\overline t}}}
\def\bbbar{\ensuremath{{b\bbar}}}
\def\ttbar{\ensuremath{{t\tbar}}}

\def\invfb{\ensuremath{{\mathrm{fb}^{-1}}}}

\def\fb{\ensuremath{{\mathrm{fb}}}}
\def\pb{\ensuremath{{\mathrm{pb}}}}
\def\nb{\ensuremath{{\mathrm{nb}}}}

\newcommand{\LUMI}{140\,\invfb }

\clearpage

\begin{document}

\maketitle

\section{Introduction}
%

\begin{figure}[bp!]
\centering
\subfloat[]{\includegraphics[width=0.3\linewidth]{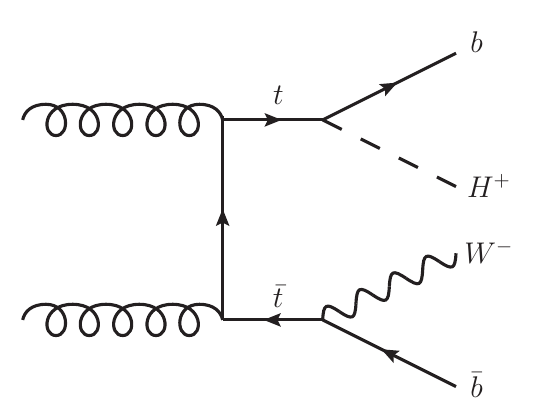}}
\subfloat[]{\includegraphics[width=0.3\linewidth]{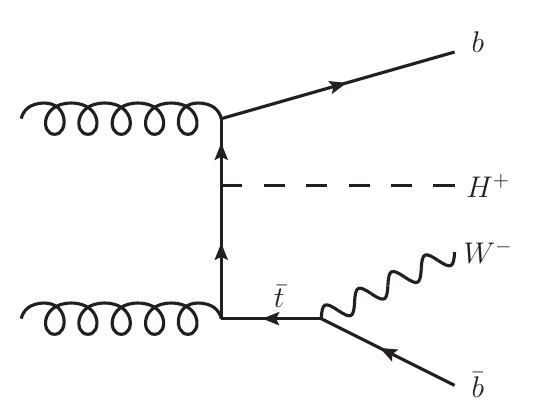}}
\subfloat[]{\includegraphics[width=0.3\linewidth]{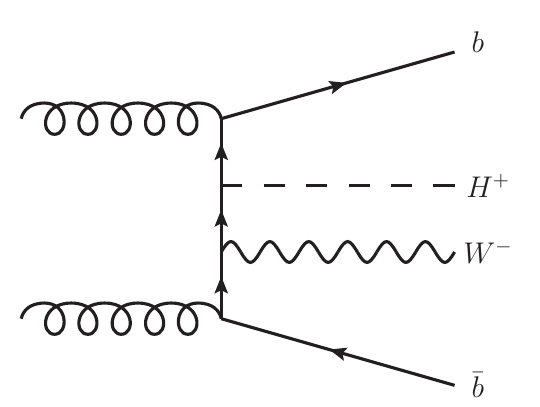}}
\caption{\label{fig:hplus-diagrams} Examples of leading-order Feynman diagrams contributing to the production of
charged Higgs bosons in $pp$ collisions:
(a) double-resonant top-quark production that dominates at low $\Hp$ masses,
(b) single-resonant top-quark production that dominates at large $\Hp$ masses,
(c) top-quark exchange production.
The interference between these three main diagrams becomes most relevant in the intermediate-mass region.}
\end{figure}

%
%
%
%
The discovery of a new boson at the Large Hadron Collider (LHC)~\cite{Evans:2008zzb}
in 2012~\cite{HIGG-2012-27,CMS-HIG-12-028}, with a measured mass
close to 125~\GeV~\cite{CMS-HIG-19-004,HIGG-2022-20,CMS:2024eka}, opens the question of whether this
particle could be part of an extended scalar sector.
Charged Higgs bosons\footnote{In the
following, charged Higgs bosons are denoted $\Hp$, with the
charge-conjugate $\Hm$ always implied. Generic symbols are also used for particles
produced in association with charged Higgs bosons and in their decays.} are predicted in several
extensions of the Standard Model (SM) that add a second doublet~\cite{Djouadi:2005gj,Branco:2011iw}
or triplets~\cite{Senjanovic:1975rk,CiezaMontalvo:2006zt,Gunion:1989ci,Georgi:1985nv}
to its scalar sector.
For \Hp\ masses below the top-quark mass ($m_{\Hp} < m_{\text{top}}$), the main production
mechanism is through the decay of a top quark, $t \to b\Hp$, in double-resonant top-quark
production ($t\bar{t}$). In this mass range, the decay $\Hp \to \tau\nu$ usually dominates in a two-Higgs-doublet model (2HDM) type-II,
although $\Hp \to cs~\mbox{and}~cb$ may also become sizable at low ratio of the vacuum expectation values of the two Higgs doublets, $\tan\beta$.
For \Hp\ masses above the top-quark mass ($m_{\Hp} > m_{\text{top}}$), the leading production mode is
$gg \to tb\Hp$ (single-resonant top-quark production)~\cite{Degrande:2015vpa}.
For the heavy \Hp, the $\Hp \to tb$ channel is dominant, but since the coupling of \Hp\ to leptons is proportional to $\tan \beta$, the branching fraction $\Hp \to \tau \nu$ remains sizable for large values of \( \tan \beta \).
In the intermediate-mass region ($m_{\Hp} \simeq m_{\text{top}}$) the \Hp\ production occurs via the double-resonant, single-resonant and top-quark exchange production processes, along with their interference~\cite{Degrande:2016hyf}.
Figure~\ref{fig:hplus-diagrams} illustrates the main production modes for charged Higgs bosons in proton-proton ($pp$) collisions.

There are many earlier searches for charged Higgs bosons at colliders.
The Large Electron-Positron Collider (LEP) experiments excluded the \Hp\ mass below 80~\GeV
at 95\% confidence level (CL), considering only the decays $\Hp \to cs$ and
$\Hp \to \tau\nu$ under the assumption $\mathrm{\cal{B}}(\Hp \to cs)$ + $\mathrm{\cal{B}}(\Hp \to \tau\nu)$ = 1~\cite{LEPComb}.
The ATLAS and CMS collaborations searched for charged Higgs bosons in Run~1 with $pp$ collisions at
$\sqrt{s}$ = 7--8~\TeV probing the mass range below the top-quark mass with the
$\tau\nu$~\cite{HIGG-2012-09,HIGG-2012-22,HIGG-2013-30,CMS-HIG-11-019,CMS-HIG-14-023,},
$cb$~\cite{CMS-HIG-16-030} and $cs$~\cite{HIGG-2012-10,CMS-HIG-13-035} decay modes,
as well as the mass range above the top-quark mass with the $\tau\nu$, $tb$, $WZ$ decay modes~\cite{HIGG-2013-30,CMS-HIG-14-023,HIGG-2013-28,HIGG-2014-13}.
Both experiments also performed searches with Run~2 $pp$ collisions at $\sqrt{s}$ = 13~\TeV in the $\tau\nu$~\cite{HIGG-2016-11,CMS-HIG-18-014}, $tb$~\cite{HDBS-2018-51,CMS-HIG-18-015}, $WZ$~\cite{EXOT-2016-11,CMS-HIG-20-017}, $HW$~\cite{CMS-WH,ATLAS-WH}, $cb$~\cite{HDBS-2019-24} and $cs$~\cite{CMS-HIG-18-021,hpluscs2024} decay channels.
The intermediate mass region was probed for the first time in the search for $\Hp \to \tau\nu$~\cite{HIGG-2016-11,CMS-HIG-18-014}.

No evidence of charged Higgs bosons was found in any of these searches.
The ATLAS and CMS experiments also searched for neutral scalar resonances
decaying to a $\tau$-lepton pair~\cite{HDBS-2018-46,CMS-HIG-21-001},
sensitive to the habemus Minimal Supersymmetric Standarm Model (hMSSM)~\cite{Djouadi:2013uqa, Djouadi:2015jea} in some regions of its parameter space.
A recent Run~2 $\Hp \to ~cb$ analysis by the ATLAS Collaboration~\cite{HDBS-2019-24} observed an excess above the SM background with a global significance of $2.5\sigma$ at the 130~\GeV $\Hp$ mass hypothesis.
This result, coupled with the tension in semi-leptonic $B$-meson decays from the combination of measurements at several flavor experiments~\cite{RDRDstar}, support the viability of the 2HDM which can accommodate flavor changing neutral currents~\cite{Type3-2hdm-off-diag,RDStar-2HDM}. Therefore, further investigation into the decays of the charged Higgs boson is well motivated~\cite{RDstar-r-parity}.

This publication describes a search for charged Higgs bosons in the mass range 80--3000~\GeV, produced either in top-quark decays or in association with a top quark, decaying via $\Hp \to \tau\nu$, with a subsequent decay of the $\tau$-lepton into a neutrino and hadrons (referred to as $\tau_{\text{had}}$).
Depending on the decay mode of the $W$ boson originating from the top quark produced together with the \Hp, two channels are targeted:
$\tau_{\text{had}}$+jets if the $W$ boson decays into a $q\bar{q}'$ pair,
or $\tau_{\text{had}}$+lepton if the $W$ boson decays into an electron
or muon and at least one neutrino (directly or via a leptonically decaying $\tau$-lepton).
The search is optimized for a generic 2HDM type-II scenario~\cite{Branco:2011iw} and results are presented as 95\% CL upper limits on cross-section times branching fraction for the $\Hp$ production and decay to the $\tau\nu$ final state.
Interpretations are performed in the context of
the hMSSM
and the $M_h^{125}$ scenario of the Minimal Supersymmetric Standard Model (MSSM) where all superpartners are chosen to be heavy~\cite{Heinemeyer:2019}.

This analysis uses the full Run~2 dataset of $pp$ collisions at $\sqrt{s}=13$~\TeV collected with the ATLAS experiment at the LHC, corresponding to \LUMI, and thus it supersedes the result of an earlier search based on the partial Run~2 dataset~\cite{HIGG-2016-11}.
With respect to the previous analysis, the mass range of the search is expanded (previously 90--2500~\GeV), the identification of $\tau_{\text{had}}$ candidates benefits from a new approach based on recurrent neural networks providing better signal purity, and the data-driven estimation of background from misidentified $\tau_{\text{had}}$ candidates is upgraded.  Additionally, the modeling of the efficiency of the missing transverse energy trigger in simulation, as well as the modeling of $t\bar{t}$ and $W$+jets backgrounds is improved.
The final signal-to-background discriminating variable is changed from a boosted decision tree~\cite{BDT} to a mass parameterized neural network (PNN)~\cite{Baldi:2016fzo}, similar to e.g. the \Hp\ search in the $tb$ channel presented in Ref.~\cite{HDBS-2018-51}.
These improvements result in an increased sensitivity to a hypothetical \Hp\ signal by up to a factor of two (four) in the low (high) mass region compared to the previous ATLAS search~\cite{HIGG-2016-11}.
%
%

%
%
%
%
%
%


%

%
\section{ATLAS detector}
%

%
%
%
%

The ATLAS experiment~\cite{PERF-2007-01} at the LHC is a multipurpose particle detector
with a forward--backward symmetric cylindrical geometry and a near \(4\pi\) coverage in
solid angle.\footnote{\AtlasCoordFootnote}
It consists of an inner tracking detector surrounded by a thin superconducting solenoid
providing a \qty{2}{\tesla} axial magnetic field, electromagnetic and hadronic calorimeters, and a muon spectrometer.
The inner tracking detector covers the pseudorapidity range \(|\eta| < 2.5\).
It consists of silicon pixel, silicon microstrip, and transition radiation tracking detectors.
Lead/liquid-argon (LAr) sampling calorimeters provide electromagnetic (EM) energy measurements
with high granularity within the region \(|\eta|< 3.2\).
A steel/scintillator-tile hadronic calorimeter covers the central pseudorapidity range (\(|\eta| < 1.7\)).
The endcap and forward regions are instrumented with LAr calorimeters
for EM and hadronic energy measurements up to \(|\eta| = 4.9\).
The muon spectrometer surrounds the calorimeters and is based on
three large superconducting air-core toroidal magnets with eight coils each.
The field integral of the toroids ranges between \num{2.0} and \qty{6.0}{\tesla\metre}
across most of the detector.
The muon spectrometer includes a system of precision tracking chambers up to \(|\eta| = 2.7\) and fast detectors for triggering up to \(|\eta| = 2.4\).
The luminosity is measured mainly by the LUCID--2~\cite{LUCID2} detector, which is located close to the beampipe.
A two-level trigger system is used to select events~\cite{TRIG-2016-01}.
The first-level trigger is implemented in hardware and uses a subset of the detector information
to accept events at a rate below \qty{100}{\kHz}.
This is followed by a software-based trigger that
reduces the accepted event rate to \qty{1}{\kHz} on average
depending on the data-taking conditions.
A software suite~\cite{SOFT-2022-02} is used in data simulation, in the reconstruction
and analysis of real and simulated data, in detector operations, and in the trigger and data acquisition
systems of the experiment.


%

%

\section{Data and simulated event samples}
%

The dataset used in this analysis, collected during stable beam conditions and with all ATLAS subsystems fully operational, corresponds to an integrated luminosity of \LUMI\ of $pp$ collisions at $\sqrt{s}=13$~\TeV collected during 2015--2018~\cite{lumi}.

The simulated Monte Carlo (MC) samples used are based on the ATLAS full \textsc{Geant4} simulation~\cite{Agostinelli:2002hh,SOFT-2010-01} and are reconstructed using the same analysis
chain as the data.
SM background samples include \ttbar\ and single-top-quark production, $W$ and $Z$ plus jets with leptonic vector boson decays and diboson production.
All samples are scaled by $k$-factors so that the estimated background yields correspond to the recent theoretical predictions on the cross-sections at next-to-next-to-leading order (NNLO) or next-to-leading order (NLO).
Finally, all MC events are overlaid with additional minimum-bias events generated with \Pythia v8.186~\cite{pythia81} using the A3 set of tuned parameters~\cite{a3tune} and the \textsc{NNPDF2.3lo}~\cite{Ball:2012cx} set of parton distributions functions (PDF) to simulate the effect of multiple $pp$ collisions per bunch crossing (pileup), at a variable rate.
The simulated events are then weighted to match the pileup distribution observed in data.
The full list of simulated SM backgrounds is presented in Table~\ref{tab:xs}.

\begin{table}[tp!]
\begin{center}
\small
\caption{\label{tab:xs}
List of SM background processes, generators utilized for matrix-element (ME) calculations, parton shower and hadronization, the PDF sets used, the cross-section to which the total expected event yield is normalized, and the order to which the background processes were calculated.
All background cross-sections are normalized to NNLO predictions,
except for diboson events, where the NLO prediction is used.
}

\begin{tabular}{l|l|l|l|l}
& ~~Generator and  &       & Cross-       & ME    \\
~~Background process & ~~parton shower & ~~PDF & section [pb] & Order \\
\hline
$\begin{array}{l} \ttbar %
\end{array}$ & $\begin{array}{l} \mbox{\textsc{Powheg-Box v2}}\mbox{~\cite{Alioli:2010xd}}~\text{and} \\
\mbox{\textsc{Pythia 8}}\mbox{~\cite{Sjostrand:2014zea}} \end{array}$ & $\begin{array}{l}\textsc{NNPDF3.0nlo}\mbox{~\cite{Ball:2014uwa}}\end{array}$ &
$832$ & NLO \\
\hline
$\begin{array}{l} \mbox{Single top quark} \\ \mbox{$t$-channel}  \end{array}$ &  &  & $217$ & NLO \\
$\begin{array}{l} \mbox{Single top quark} \\ \mbox{$s$-channel}  \end{array}$ &
$\begin{array}{l} \mbox{\textsc{Powheg-Box v2}}~\text{and} \\
\mbox{\textsc{Pythia 8}} \end{array}$ &
$\begin{array}{l}\textsc{NNPDF3.0nlo}\end{array}$  &
$10.3$ &
NLO \\
$\begin{array}{l} \mbox{Single top quark} \\ \mbox{$Wt$-channel} \end{array}$ &  &  & $72$ & NLO \\
\hline
$\begin{array}{l} W(\ell\nu) + \mbox{jets} \end{array}$ &$\begin{array}{l} \mbox{\textsc{Sherpa 2.2.1}}\mbox{~\cite{Bothmann:2019yzt}}\end{array}$ & $\begin{array}{l}\textsc{NNPDF3.0nnlo}\end{array}$ &
$2.0\times 10^4$ &
NNLO \\
\hline
$\begin{array}{l} Z/\gamma^{\ast}(\ell\ell,\nu\nu) + \mbox{jets}\end{array}$ &
$\begin{array}{l}\mbox{\textsc{Sherpa 2.2.1}}\end{array}$ &
$\begin{array}{l}\textsc{NNPDF3.0nnlo}\end{array}$ &
$2.1 \times 10^3$ &
NNLO \\
\hline
$\begin{array}{l}  \mbox{$WW$}\end{array}$  &  &  & $55$ & NLO \\
$\begin{array}{l}  \mbox{$WZ$}\end{array}$  & $\begin{array}{l} \mbox{\textsc{Powheg-Box v2}}~\text{and} \\ \mbox{\textsc{Pythia 8}} \end{array}$ &$\begin{array}{l} \mbox{\textsc{CT10nlo}}\mbox{~\cite{Lai:2010vv}} \end{array}$ & 26 & NLO \\
$\begin{array}{l}  \mbox{$ZZ$}\end{array}$  &  &  & $8.4$ & NLO \\
\end{tabular}
\end{center}
\end{table}

Simulated events of $\Hp$ signal are generated in three distinct mass regions using the narrow-width approximation with \MGNLO~\cite{Alwall:2014hca} at either leading order (LO) or NLO in QCD using the respective \textsc{NNPDF3.0} set.
The choice of renormalization and factorization scale settings is motivated by Refs.~\cite{Degrande:2016hyf,Degrande:2015vpa}.
\begin{enumerate}
\item{
In the low mass region ($m_{H^+}<140~\mathrm{GeV}$), \ttbar\ events with a subsequent decay of one top quark to a $\Hp$ and a bottom quark are generated
using LO calculations only.
The type-II 2HDM model is used and dynamic renormalization and factorization QCD scales are chosen
($\mu_{\mathrm{R}}=\mu_{\mathrm{F}}=H_\mathrm{T}/2=1/2~\sum_i{\sqrt{m_i^2+p_{\mathrm{T}i}^2}}$, where the sum goes over all final state partons from the hard scatter).
The contribution from \ttbar\ events with both top quarks decaying to $b \Hp$ and single-top-quark events with a subsequent $t\rightarrow b \Hp$ decay is negligible.
}
\item{
In the intermediate mass region ($140~\mathrm{GeV} \le m_{H^+}<200~\mathrm{GeV}$), non-resonant, single- and double-resonant top-quark processes with a $W$ boson, a $\Hp$ and two bottom quarks in the final state are generated in the four-flavor scheme (4FS)\footnote{In the matrix-element calculation, only gluons and first- and second-generation quarks are considered when defining the proton parton distribution functions.}
at LO.
The type-II 2HDM model
with static QCD scales ($\mu_\mathrm{R}=\mu_\mathrm{F}=125~\mathrm{GeV}$) is used.
}
\item{
In the high mass region ($m_{H^+} \ge 200~\mathrm{GeV}$), $\Hp$ production in association with a single top quark is generated in the 4FS at NLO.
The type-II 2HDM model is used and the dynamic QCD scales are chosen
($\mu_\mathrm{R}=\mu_\mathrm{F}=H_\mathrm{T}/3=1/3~\sum_i{\sqrt{m_i^2+p_{\mathrm{T}i}^2}}$).
}
\end{enumerate}
For all signal samples,
the parton-level generator is interfaced to \Pythia8~\cite{Sjostrand:2014zea} with the \textsc{NNPDF2.3lo} PDF set
and the A14~\cite{ATL-PHYS-PUB-2014-021} set of tuned parameters.


%
\section{Analysis}
%

The general analysis strategy, objects definition, event selection as well as background and signal modeling follow those of Ref.~\cite{HIGG-2016-11}.

\subsection{Event reconstruction and selection}
\label{subsec:event_selection} %

Recorded events are filtered by requiring at least one primary vertex~\cite{ATL-PHYS-PUB-2015-026} with two or more associated tracks with $\pt > 400~\MeV$, that they pass a good-quality requirement, and that all relevant detector components were in good operating condition~\cite{DAPR-2018-01}.

Hadronic $\tau$-lepton~\cite{ATL-PHYS-PUB-2015-045,ATL-PHYS-PUB-2019-033} decays are identified using a Recurrent Neural Network (RNN) designed to discriminate
against quark- and gluon-initiated jets~\cite{ATLAS-CONF-2017-029}.  %
In the following, reconstructed $\tau$ candidates, corresponding to the visible part of a hadronic $\tau$-lepton decay (hereafter called $\tauhadvis$), are required to have $\pt > 20~\GeV$, $|\eta| < 2.5$, excluding $1.37 < |\eta| < 1.52$, an electric charge of +1 or -1, and one or three associated tracks, also referred to as 1- or 3-prong, respectively. A separate boosted decision tree is used to reject electrons that are misidentified as 1-prong $\tauhadvis$ candidates.
Electrons and muons are reconstructed and identified as reported in Refs.~\cite{EGAM-2021-01} and~\cite{MUON-2018-03}, respectively.
Jets are reconstructed using the particle flow approach~\cite{PERF-2015-09} and clustered with the anti-$k_t$ algorithm~\cite{Cacciari:2008gp,Fastjet} with a radius parameter $R=0.4$.
Additionally, jets originating from the hadronization of bottom quarks are identified ($b$-tagged) using advanced algorithms combined into the DL1r tagger as described in Ref.~\cite{FTAG-2019-07}.
A working point corresponding to an average efficiency of 70\%\ for $b$-jets in simulated $t\bar{t}$ events is chosen.
When several objects defined above overlap geometrically, an overlap removal procedure is applied as described in Ref.~\cite{HIGG-2016-11}.
The missing transverse momentum in the event, with magnitude $\met$, is determined from the reconstructed objects according to Ref.~\cite{JETM-2020-03}.
Simulated events are corrected for differences between data and MC simulation seen in $b$-tagging efficiencies and mis-tag rates as well as minor differences in electron ($e$), muon ($\mu$) and $\tauhadvis$ reconstruction, identification, and isolation efficiencies.

The analysis of the $\tau_{\text{had}}$+jets channel is based on events accepted by $\met$ triggers~\cite{TRIG-2019-01} with a threshold at 70, 80, 90 or 110~\GeV, depending on the data-taking period.\footnote{Due to differing pile-up conditions which impact the trigger rates.}
Because the \met\ trigger is not accurately modeled in simulation, the efficiency of the trigger decision as a function of reconstructed \met\ is fitted using a Gaussian error function ($erf (x) = \frac{1}{\sqrt{2\pi\sigma^2}}\int_{-\infty}^x e^{-(t-\mu)^2/2\sigma^2} dt$) in both data and simulation, following a method similar to that in Ref.~\cite{HIGG-2015-11}. The ratio of the \emph{erf} fit in data to that in simulation is then used to reweight the simulated events.

The $\tau_{\text{had}}$+jets channel signal region (SR) further requires at least one \emph{medium} $\tauhadvis$ candidate, corresponding to 75\%\ (60\%) efficiency for 1-prong (3-prong) $\tauhadvis$ candidates~\cite{ATL-PHYS-PUB-2019-033}, with $\pt > 40~\GeV$ and $|\eta| < 2.3$, no \emph{loose}
leptons (electron~\cite{EGAM-2021-01} or muon~\cite{MUON-2018-03}) with $\pt > 20~\GeV$, at least three jets with $\pt > 25~\GeV$, of which at least one is $b$-tagged, $\met > 150~\GeV$ and $\mt > 50~\GeV$.\footnote{$\mt$, the transverse mass of the highest-$\pt$ $\tauhadvis$ candidate and $\met$ is defined as $\mt(\tauhadvis,\met)=\sqrt{ 2 \pt^\tau  \met (1-\cos \Delta\phi_{\tau,\text{miss}}) }$, where $\Delta\phi_{\tau,\text{miss}}$ is the azimuthal angle between the $\tauhadvis$ candidate and the direction of the missing transverse momentum.}
A signal-depleted control region (CR) is defined for the $\tau_{\text{had}}$+jets channel ($\tau_{\text{had}}$+jets $t\overline{t}$ CR), to probe the modeling of the \ttbar\  background.
The $t\bar{t}$-enriched $\tau_{\text{had}}$+jets $t\overline{t}$ CR has the same event selection as described above, except $\mt > 50~\GeV$ is replaced with $\mt < 100~\GeV$ and at least two $b$ jets are required.
The purity of this region in top-quark backgrounds, i.e. \ttbar\ and single-top-quark events is around 90\%.
To ensure full orthogonality, events satisfying the $\tau_{\text{had}}$+jets $t\overline{t}$ CR selection criteria
are excluded from the $\tau_{\text{had}}$+jets SR.

The $\tau_{\text{had}}$+lepton channel, made of the $\tau_{\text{had}}$+electron and $\tau_{\text{had}}$+muon sub-channels, is based on events accepted
by single-lepton triggers. Triggers for electrons or muons~\cite{TRIG-2018-05,TRIG-2018-01,TRIG-2019-03} with low $E_{\text{T}}$ or $\pt$
thresholds respectively (20--26~\GeV\ depending on the data-taking period) and isolation requirements are combined in a logical OR with triggers having higher $\pt$ thresholds (60--120~\GeV\ for electrons, 50~\GeV\ for muons) and looser isolation or identification requirements to maximize the efficiency.

The $\tau_{\text{had}}$+lepton channel SR events are selected requiring exactly one \emph{tight}
lepton (electron~\cite{EGAM-2021-01} or muon~\cite{MUON-2018-03}) with $\pt > 30~\GeV$ and $|\eta| < 2.5$ ($|\eta| < 2.47$, excluding $1.37<|\eta|<1.52$) for muons (electrons) matched to the single-lepton trigger object, exactly one \emph{medium} $\tauhadvis$ candidate with $\pt > 30~\GeV$, $|\eta| < 2.3$ and an electric charge opposite to that of the lepton, at least one $b$-tagged jet with $\pt > 25~\GeV$ and $\met > 50~\GeV$.
Additionally, a $t\bar{t}$-enriched CR is defined for the $\tau_{\text{had}}$+lepton channel ($\tau_{\text{had}}$+lepton $t\overline{t}$ CR), to derive \ttbar\ modeling corrections in this channel.
This CR has the same event selection as described above, with the addition of  $\met > 80~\GeV$, $\mt < 70~\GeV$ and at least two $b$-tagged jets.
This CR selection is a subset of $\tau_{\text{had}}$+lepton selection described above.
To maintain orthogonality of the regions, events satisfying the $\tau_{\text{had}}$+lepton $t\overline{t}$ selection are rejected from the $\tau_{\text{had}}$+lepton SR.

Two further signal-depleted CRs are defined for the
$\tau_{\text{had}}$+lepton channel.
The $\tau_{\text{had}}+e/\mu$~b-veto CR relies on the same event selection as the SR, vetoing any $b$-tagged jets.
This region is enriched in $W$+jets events with true $e/\mu$ and a jet misidentified as \tauhadvis and $Z$+jets events with $Z$ boson decaying to a pair of $\tau$ leptons.
The  $t\bar{t}$ enriched CR ($e+\mu$~$b$-tag CR) uses SR event selection,
except that a \emph{tight} (and oppositely charged) $e+\mu$ pair is required instead of a \tauhadvis+lepton pair.
This CR has a very high purity ($>99.8\%$) of top-quark background events.
Control regions are used to validate the agreement between data and background simulation or derive necessary corrections, but are not included in the final likelihood fit. A summary of the selection applied to the regions used in this analysis is shown in Table~\ref{tab:regions}.

\begin{table}[tp!]
\begin{center}
\caption{Summary of requirements for the regions used in the analysis. When applicable, units are in \GeV, $\tau$ refers to $\tauhadvis$ in the cut definitions, and $j$ refers to jets in the region names.
FF MJ and FF~$W$+$j$ denote multi-jet and $W$+jets CRs, respectively, defined to determine $\tauhadvis$ fake factors as described in Section~\ref{subsec:bkg_modeling}.}
\begin{tabular}{l|r|r|r|r|r|r|r|r|r}
Cut                      & SR $\tau$+$j$    & $t\bar{t}(\tau j)$ & $W$+$j$  & FF~MJ        & FF~$W$+$j$       & SR $\tau+\ell$ & $t\bar{t}(\tau\ell)$ & $b$-veto & $e+\mu$  \\
\hline
$\pT(\tau)$              & $>40$            & $>40$              & $>40$    & $>30$        & $>30$            & $>30$            & $>30$                & $>30$    & ...      \\
$N_\ell$                 & 0                & 0                  & 0        & 0            & 1                & 1                & 1                    & 1        & 2        \\
$\pT(\ell)$              & ...              & ...                & ...      & ...          & $>30$            & $>30$            & $>30$                & $>30$    & $>30$    \\
$q(\tau) \times q(\ell)$ & ...              & ...                & ...      & ...          & ...              & -1               & -1                   & -1       & ...      \\
$N_\text{jet}$           & $\geq 3$         & $\geq 3$           & $\geq 3$ & $\geq 3$     & ...              & $\geq 1$         & $\geq 2$             & $\geq 1$ & $\geq 1$ \\
$\pT(\text{lead-jet})$   & $>25$            & ...                & $>25$    & $>25$        & ...              & $>25$            & $>25$                & $>25$    & $>25$    \\
$N_{b\text{-jet}}$       & $\geq 1$         & $\geq 2$           & 0        & 0            & 0                & $\geq 1$         & $\geq 2$             & 0        & $\geq 1$ \\
$\MET$                   & $>150$           & $>150$             & $>150$   & $<80$        & ...              & $>50$            & $>80$                & $>50$    & $>50$    \\
$m_{T}(\tau,\MET)$       & $>50$            & $<100$             & $<100$   & $>50$        & ...              & ...              & $<70$                & ...      & ...      \\
Other                    & \footnotesize{a} &                    &          &              & \footnotesize{b} & \footnotesize{a} &                      &          & \footnotesize{c} \\
\end{tabular}
\caption*{
\\
$^{\text{\footnotesize{a}}}$ Note that the SRs reject events in the overlap with the corresponding $t\bar{t}$ regions. \\
$^{\text{\footnotesize{b}}}$ There is an additional requirement that $60~\GeV < m_{T}(e,\MET) < 160~\GeV$. \\
$^{\text{\footnotesize{c}}}$ The two $\ell$ must be an oppositely charged $e+\mu$ pair.
}
\label{tab:regions}
\end{center}
\end{table}

\subsection{Background modeling}
\label{subsec:bkg_modeling}

Background modeling relies on simulated MC events for SM backgrounds containing the $\tauhadvis$ object matched to a true hadronic $\tau$-lepton decay at the generator level or containing electrons or muons reconstructed and identified as $\tauhadvis$ objects ($\ell\rightarrow\tauhad^\text{fake}$).

Events that contain quark- or gluon-initiated jets and no true \tauhad\ can enter the SR when one of these jets is reconstructed and misidentified as a \tauhadvis candidate (fake \tauhadvis).
A data-driven fake-factor (FF) method is used to estimate this background.
The ratios of the number of events with fake \tauhadvis\ objects satisfying the $\tauhadvis$ identification criteria ($N_\mathrm{CR~fake}^{\tau \mathrm{-ID}}$) to the number of events with fake anti-$\tauhadvis$ objects satisfying the very loose requirement on the \tauhadvis\ identification but failing to meet the loose identification requirement ($N_\mathrm{CR~fake}^{\mathrm{anti-}\tau\mathrm{-ID}}$)~\cite{ATL-PHYS-PUB-2019-033} are measured in the dedicated CRs.
The fake factors ($\mathrm{FF} = N_\mathrm{CR~fake}^{\tau \mathrm{-ID}} / N_\mathrm{CR~fake}^{\mathrm{anti-}\tau\mathrm{-ID}}$) are determined in bins
of $\tauhadvis$ \pt, separately for 1-prong and 3-prong candidates.
To estimate the number of background events with fake $\tauhadvis$ in the SR, the measured FFs are applied to the number of events with fake anti-$\tauhadvis$ objects obtained by requiring the SR selection, but replacing the $\tauhadvis$ with the anti-$\tauhadvis$ selection criteria
($N_\mathrm{SR~fake}^{\tau} = N_\mathrm{SR~fake}^{\mathrm{anti}-\tau \mathrm{-ID}}\times \mathrm{FF}$).
The number of events with a fake \tauhadvis\ candidate in each region is obtained by subtracting the number of simulated events containing a true \tauhad\ or $\ell\rightarrow\tauhad^\text{fakes}$ from the number of data events ($N_\mathrm{fake}^{\tau} = N_\mathrm{data}^{\tau} - N_\mathrm{MC~true}^{\tau} - N_\mathrm{MC}^{\ell\rightarrow\tau}$).

Two CRs with different fractions of quark- and gluon-initiated jets are used to determine the FFs.
The FF multi-jet CR relies on multi-jet event triggers~\cite{TRIG-2012-01} and is defined by selecting events satisfying an offline selection similar to that of the $\tau_{\text{had}}$+jets SR, but requiring $\pT(\tau)>30~\GeV$, $\met<80~\GeV$ and vetoing any $b$-tagged jets.
The FF $W$+jets CR is defined by selecting events passing an offline selection similar to that of the $\tau_{\text{had}}$+lepton SR, but requiring $60~\GeV < \mt(\ell,\met) < 160~\GeV$ and vetoing $b$-tagged jets.
In the FF multi-jet CR, the contamination arising from correctly reconstructed and identified
$\tauhadvis$ objects is small ($8\%$), while in the FF $W$+jets CR, this contamination is about $20\%$.
In the corresponding $\mathrm{anti-}\tau\mathrm{-ID}$ regions, the fraction of events with a true \tauhad\ is negligible in the FF multi-jet CR and is around $6\%$ in the FF $W$+jets CR. The contribution from true \tauhad\ events in the above mentioned categories is estimated using simulation and is subtracted from the number of observed events in each region.
The FF multi-jet CR contains similar fractions of quark- and gluon-initiated jets,
while the FF $W$+jets CR is dominated by quark-initiated jets.  The SR is expected
to have contributions from both types, but with different fractions. In order to
estimate the fraction of gluon-initiated jets in the signal region, a template-fit
method in the $\mathrm{anti-}\tau\mathrm{-ID}$ region is used, based on a discriminating variable that is sensitive to the source
of the jet. This fraction is used to estimate combined fake factors, which
are linear combinations of the fake factors measured in the multi-jet and $W$+jets CRs.
The variable chosen for the templates is the \tauhadvis\ width
defined as $w_\tau=\frac{\sum \left[ p_\mathrm{T}^\mathrm{track}\times\Delta R\left(\tauhadvis,\mathrm{track}\right) \right]}{\sum p_\mathrm{T}^\mathrm{track}}$ for tracks satisfying $\Delta R\left(\tauhadvis,\mathrm{track}\right)<0.4$.

Due to inaccurate data modeling at high jet multiplicities arising in the $t\bar{t}$ and $W$+jets simulations, data-based corrections are applied to the MC prediction, following the example of Refs.~\cite{HDBS-2018-51,HDBS-2019-24}.
Reweighting factors, $R(x)$, are derived by comparing the MC prediction to the data in dedicated CRs, separately for
$t\bar{t}$ and $W$+jets MC samples.
For the $\tau_{\text{had}}$+lepton channel, $R(x)$ is determined in the $\tau_{\text{had}}$+lepton $t\overline{t}$ CR as a function of $m_{\textrm{eff}}=\sum_{\textrm{jets}} p_\mathrm{T}^{\textrm{jets}} + p_\mathrm{T}^{\tau} + p_\mathrm{T}^{\textrm{lepton}}$ + $\met$
and for the $\tau_{\text{had}}$+jets channel in the $\tau_{\text{had}}$+jets $t\overline{t}$ CR as a function of $m_{\textrm{eff}}=\sum_{\textrm{jets}} p_\mathrm{T}^{\textrm{jets}} + p_\mathrm{T}^{\tau}$ + $\met$.
Reweighting factors for the $W$+jets sample are derived as a function of jet multiplicity in the $W$+jets CR of the $\tau_{\text{had}}$+jets channel and are applied to the simulated $W$+jets events in both $\tau_{\text{had}}$+jets and $\tau_{\text{had}}+$lepton channels.
The overall event normalization factor from applying the reweighting factors ranges from $0.86$ to $0.95$.

\subsection{Multivariate discriminant}

Following the event selections described in Section~\ref{subsec:event_selection}, kinematic variables that differentiate between the signal and backgrounds are combined into a multivariate signal-to-background discriminant using PNNs.
The PNN response is parameterized with the generator-level \Hp\ mass, $m_{\text{truth}}^{H^+}$, and defines classifiers for all probed $\Hp$ mass hypotheses, with continuous sensitivity to masses between them.
The PNN output score for a given $\Hp$ mass hypothesis discriminates between that signal and the SM backgrounds, and is used as the final discriminating variable for the statistical analysis.
The training of the PNNs is performed using the \textsc{Keras}~\cite{chollet2015keras} library with the \textsc{TensorFlow}~\cite{tensorflow2015-whitepaper} library as a backend.
Two separate PNNs are defined and trained for the $\tau_{\text{had}}$+jets and $\tau_{\text{had}}$+lepton channels using all signal samples at the same time, taking the value of the \Hp\ mass as a parameter.\footnote{In the training process, samples for all $\Hp$ mass hypotheses are normalized to the weighted number of background events.
Background events enter the training multiple times, such that each event is assigned to every $\Hp$ mass hypothesis once.}
The k-fold training method~\cite{stone1974cross} is used with k = 5 in order to increase the effective statistics of the training sample and to prevent overtraining.

The input variables used to train the PNNs for the $\tau_{\text{had}}$+jets and $\tau_{\text{had}}$+lepton channels are mostly the four-momentum components of the reconstructed final state objects and are listed in Table~\ref{Tab:pnn}.
For events with more than one \tauhadvis\ candidate or more than one $b$-tagged jet, the highest-$\pt$ object is passed as input to the PNN.
When comparing the PNN response for data and for the background model in regions that do not contain a \tauhadvis or $b$ jets, a default value of zero is used for the corresponding input variables.
The $\Upsilon = 2 (\pt^{\tau{\text{-track}}} / \pt^{\tauhadvis})-1$ variable~\cite{STDM-2011-46}, where $\pt^{\tau{\text{-track}}}$ refers to the track associated with the $\tauhadvis$, is sensitive to $\tau$-lepton polarization and helps discriminate between the SM background processes, where the \tauhadvis\ object stems from a vector-boson decay and signal, where \tauhad\ is a decay product of a scalar particle.
This variable is defined only for 1-prong tau decays, hence included uniquely in the PNN variable for events with a selected 1-prong \tauhadvis\ object.
Due to correlations with some input variables entering the RNN $\tauhadvis$ identification, the $\Upsilon$ distribution significantly differs between $\tauhadvis$ and anti-$\tauhadvis$ candidates.
In order to ensure proper modeling of this distribution, an inverse transform sampling (Smirnov transform) is performed on the $\Upsilon$ variable of anti-$\tauhadvis$ candidates entering the fake-factor background estimation~\cite{HIGG-2016-11}.

\begin{figure}[t]
\centering
\subfloat[]{\includegraphics[width=0.45\textwidth]{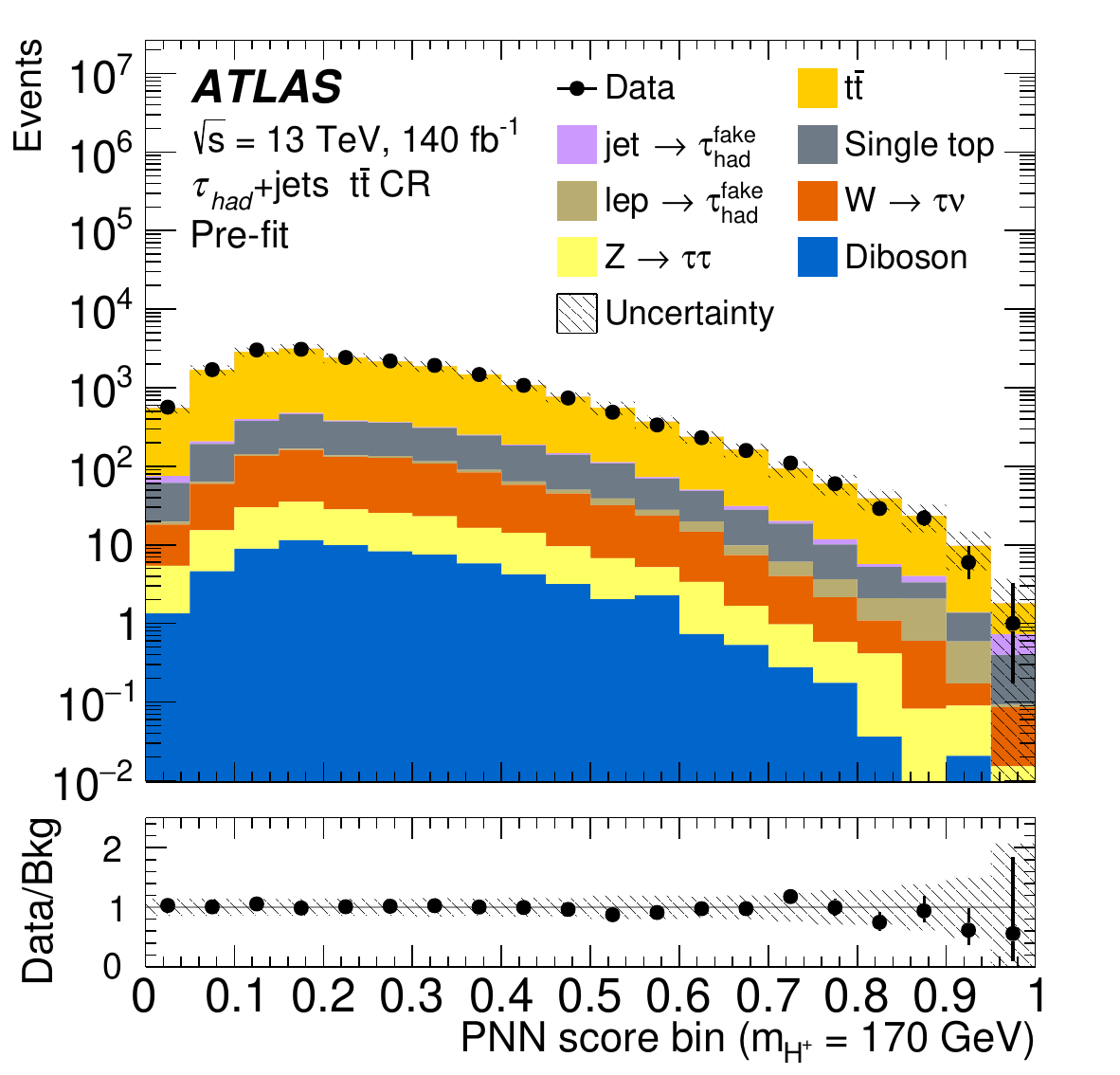}}  \qquad
\subfloat[]{\includegraphics[width=0.45\textwidth]{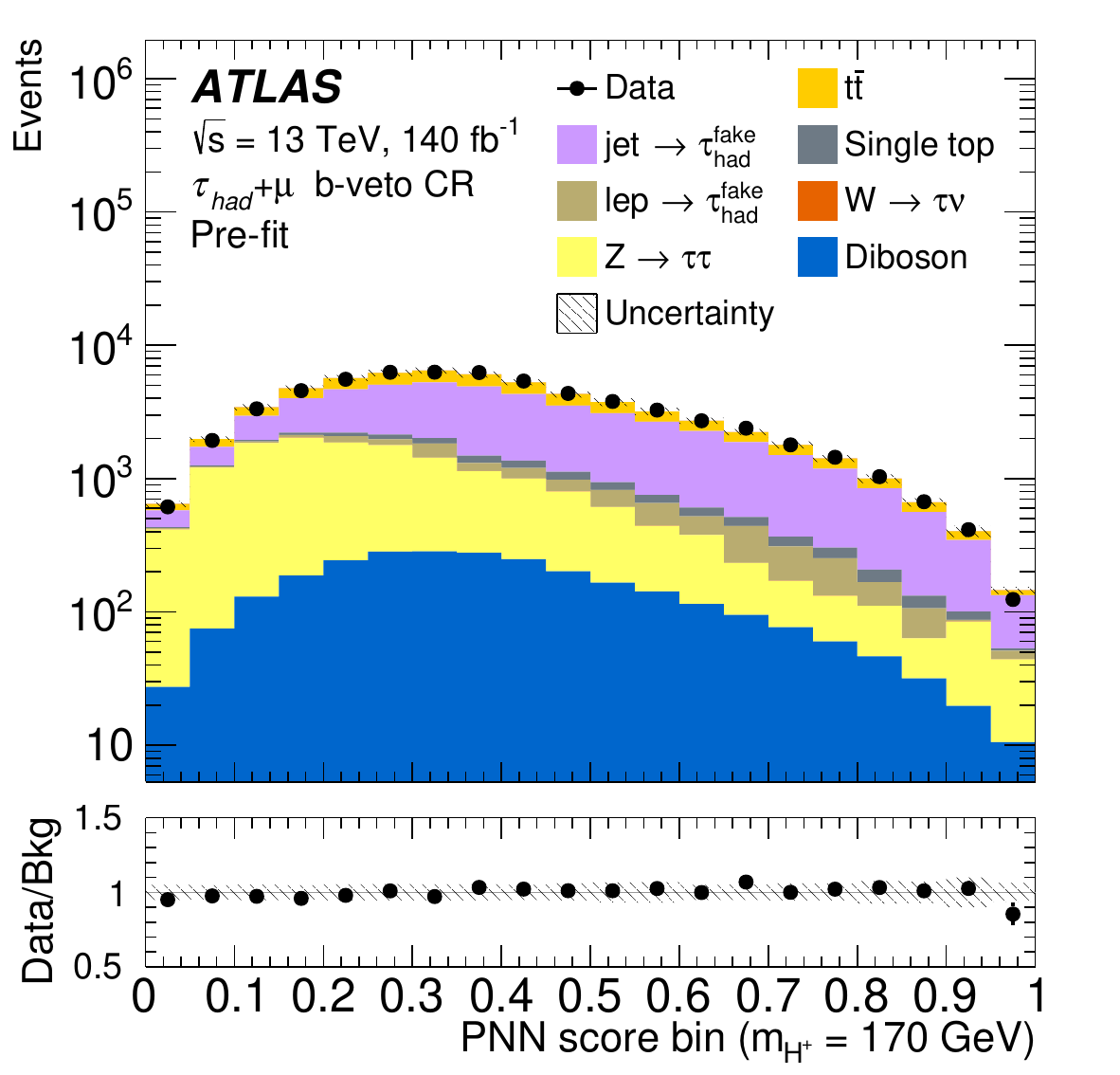}}
\caption{\label{fig:CR_PNNscores} Predicted and measured PNN score distributions before the fit to data (``Pre-fit'') in (a) the $t\bar{t}$ enriched CR of the $\tau_{\text{had}}$+jets channel and (b) the $b$-veto CR of the $\tau_{\text{had}}$+muon channel enriched in the $W/Z$+jets events with jets faking \tauhadvis, for the 170 GeV $m_{\Hp}$ hypothesis.
The lower panel of each plot shows the ratio of data to the SM background prediction.
The uncertainty bands include all statistical and systematic uncertainties.
The bin boundaries of these distributions are spaced equally at intervals of 0.05 and do not represent the binning scheme used in the final fits.
}
\end{figure}

\begin{figure}[t]
\centering
\subfloat[]{\includegraphics[width=0.45\textwidth]{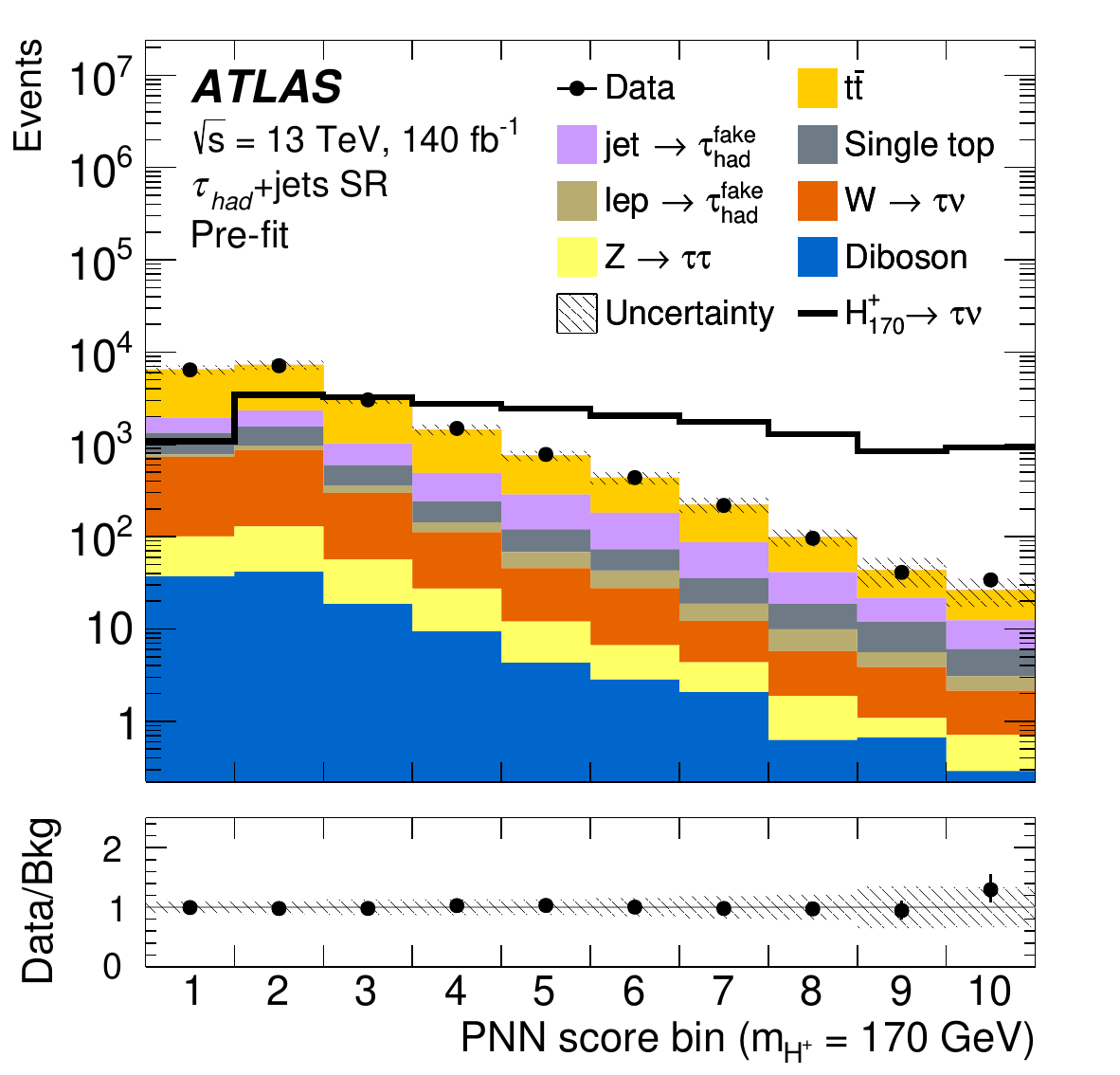}}
\subfloat[]{\includegraphics[width=0.45\textwidth]{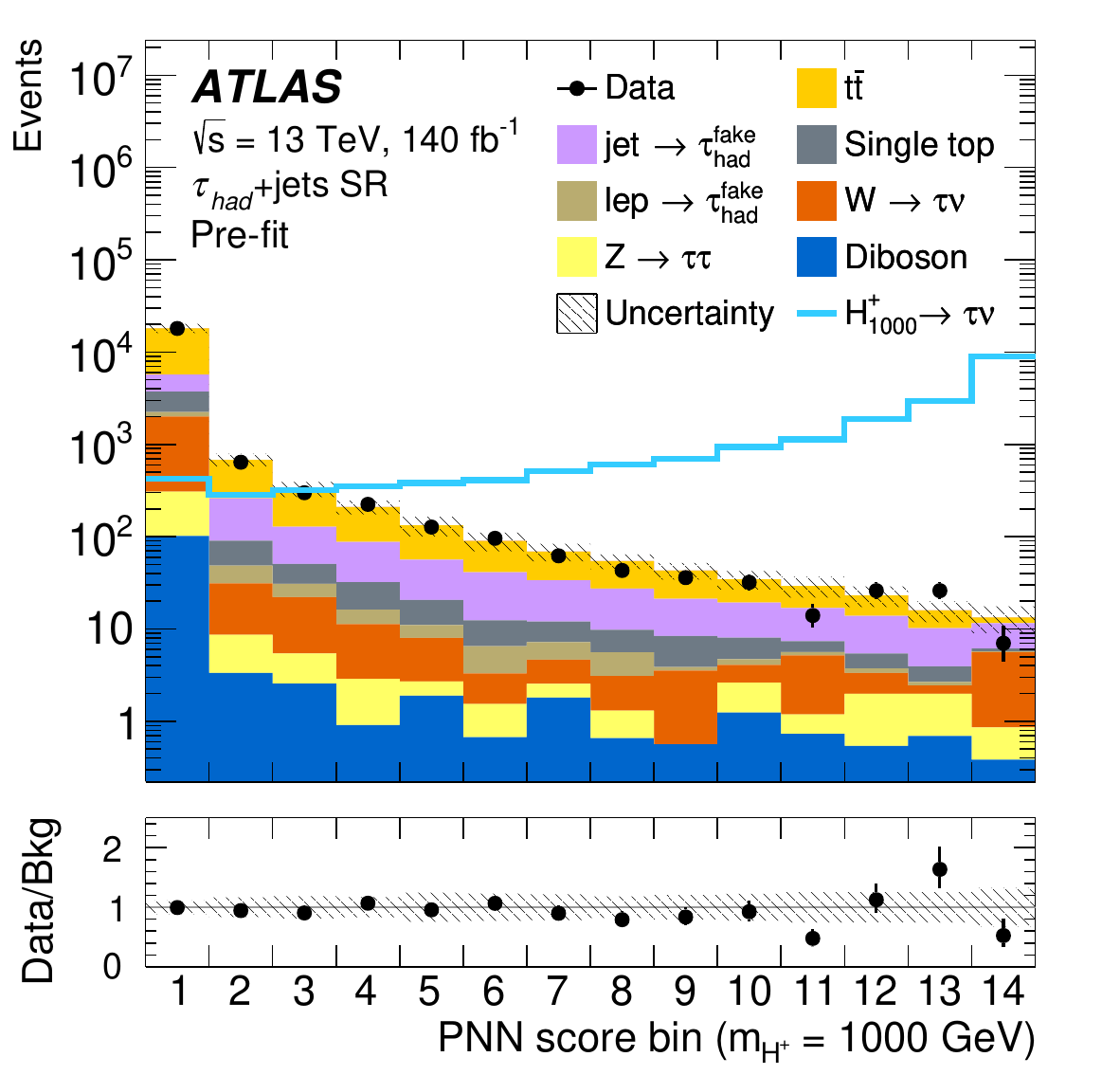}}

\caption{\label{fig:SRjets_PNNscores} PNN score distributions before the fit to data (``Pre-fit'') in the SR of the $\tau_{\text{had}}$+jets for the (a) 170~\GeV\ and (b) 1000~\GeV\ $m_{\Hp}$ hypotheses.
The lower panel of each plot shows the ratio of data to the SM background prediction.
The uncertainty bands include all statistical and systematic uncertainties.
Bin boundaries are optimized for the best sensitivity of the final fit (see Section~\ref{sec:result}) but are displayed as equidistant for better readability.
Also shown are the expected signal distributions normalized to the total background yield.
}
\end{figure}

\begin{figure}[t]
\centering
\subfloat[]{\includegraphics[width=0.45\textwidth]{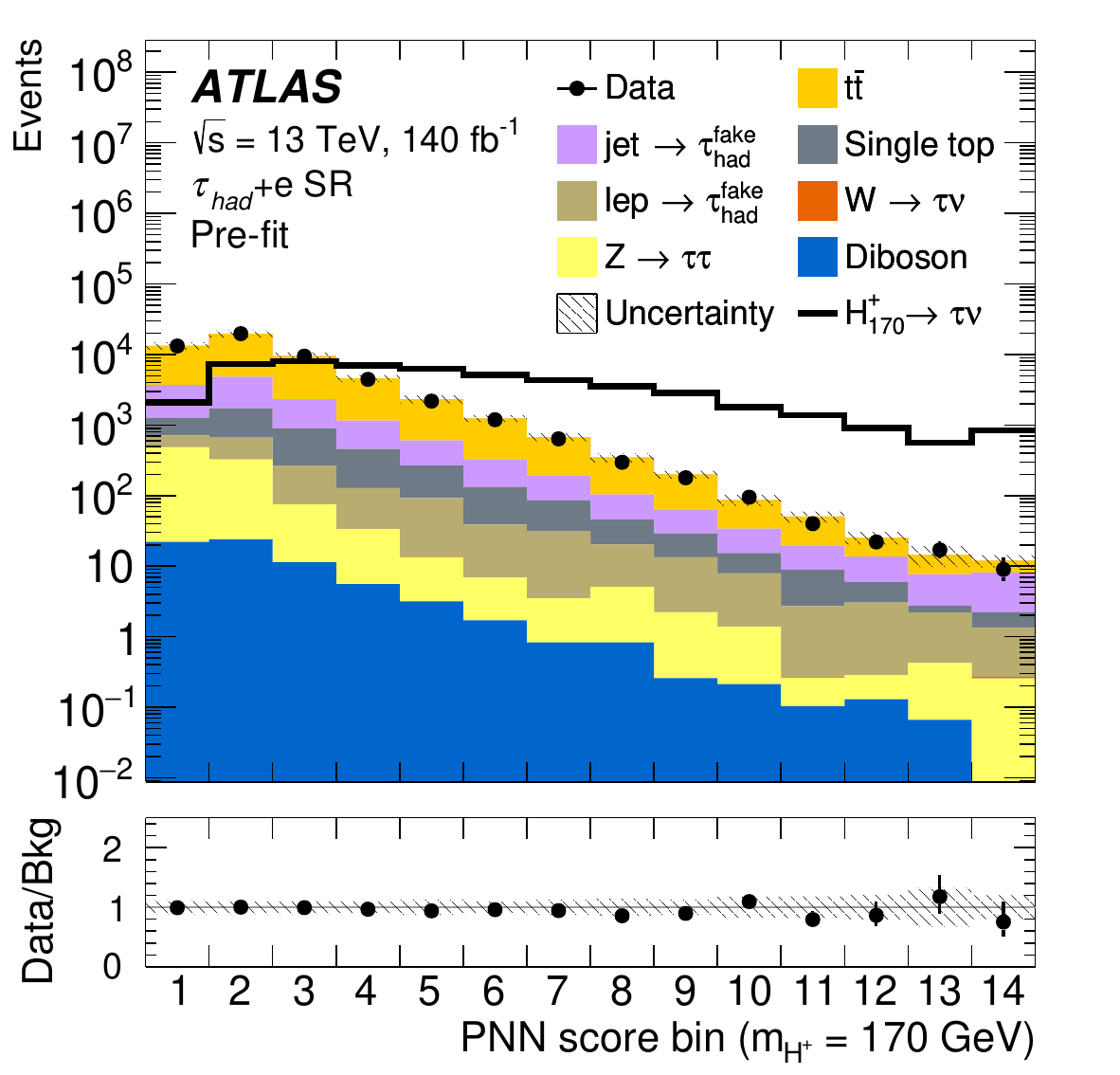}}
\subfloat[]{\includegraphics[width=0.45\textwidth]{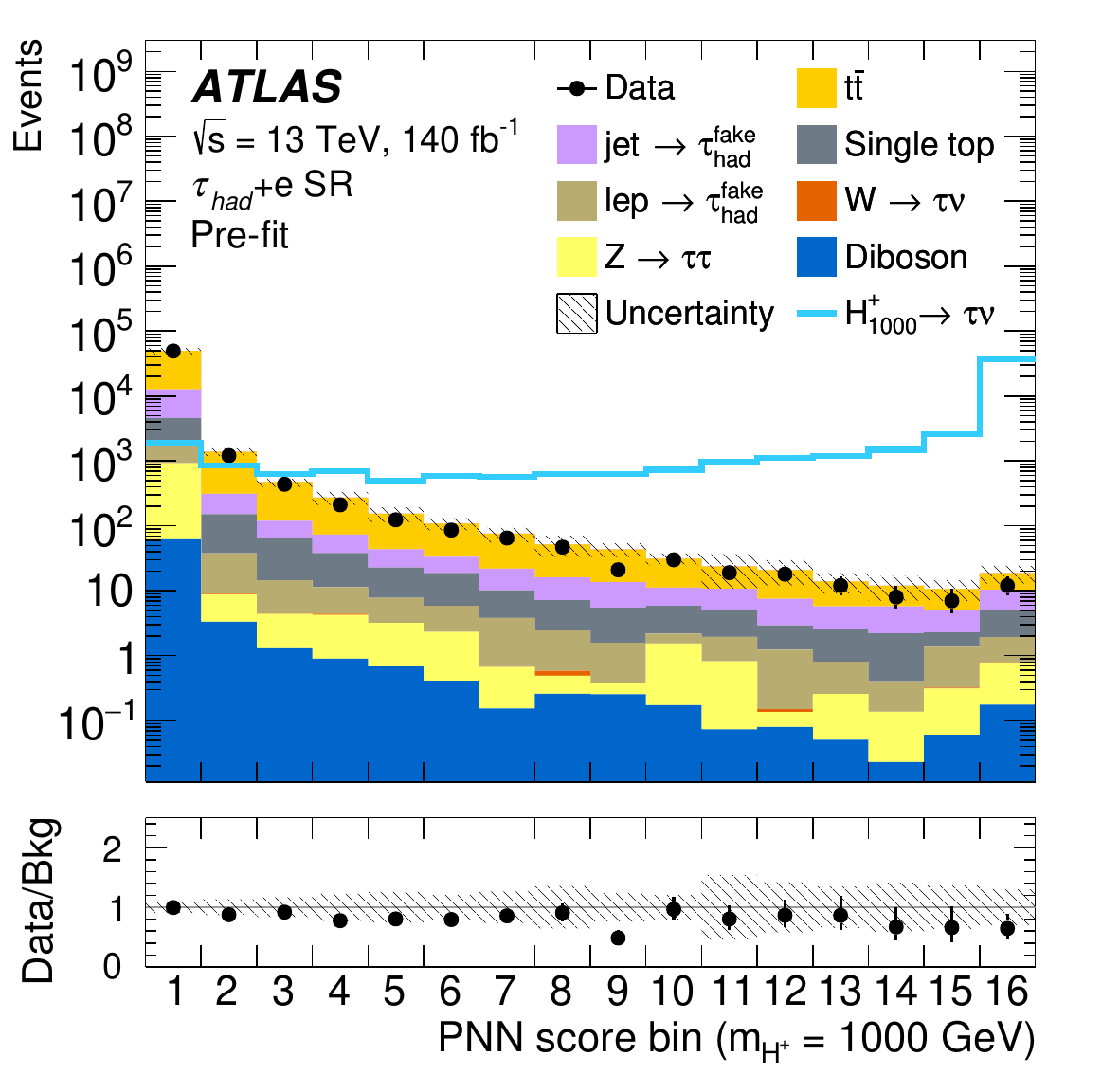}}  \\
\subfloat[]{\includegraphics[width=0.45\textwidth]{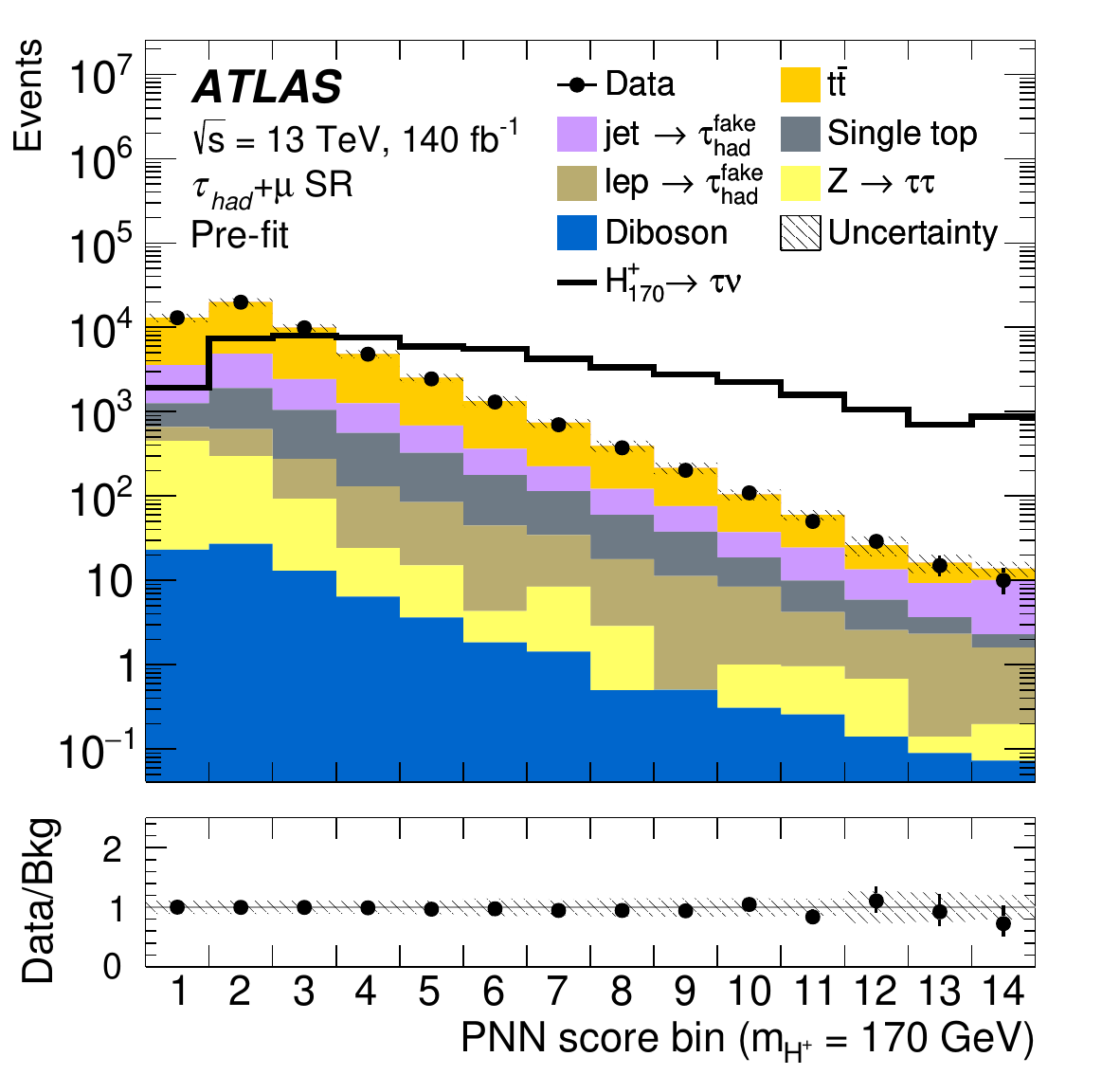}}
\subfloat[]{\includegraphics[width=0.45\textwidth]{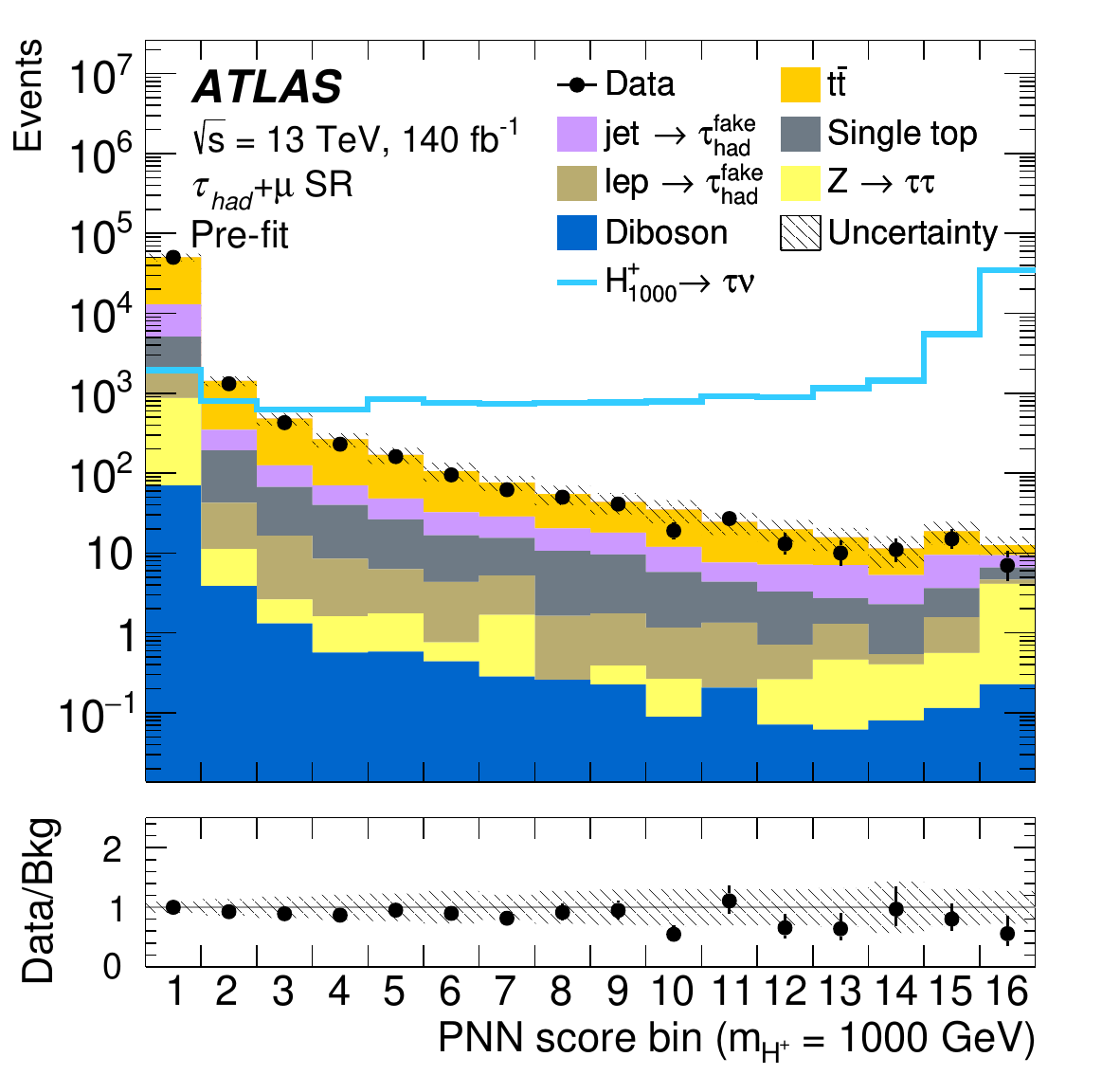}}

\caption{\label{fig:SRlep_PNNscores} PNN score distributions before the fit to data (``Pre-fit'') in the SR of the $\tau_{\text{had}}$+electron sub-channel for the (a) 170~\GeV\ and (b) 1000~\GeV\ $m_{\Hp}$ hypotheses and $\tau_{\text{had}}$+muon sub-channel for the (c) 170~\GeV\ and (d) 1000~\GeV\ $m_{\Hp}$ hypotheses.
The lower panel of each plot shows the ratio of data to the SM background prediction.
The uncertainty bands include all statistical and systematic uncertainties.
Bin boundaries are optimized for the best sensitivity of the final fit (see Section~\ref{sec:result}) but are displayed as equidistant for better readability.
Also shown are the expected signal distributions normalized to the total background yield.
}
\end{figure}

The modeling of the PNN input variables as well as the output score is validated in the CRs.
The PNN score in the \ttbar\ enriched $\tau_{\text{had}}$+jets $t\overline{t}$ CR, and the $W/Z$+jets enriched b-veto CR of the $\tau_{\text{had}}+\mu$ channel are shown in Figure~\ref{fig:CR_PNNscores} for an example \Hp\ mass of 170~\GeV .
The %
PNN score distributions for two example signal masses (170~\GeV and 1000~\GeV) are shown in Figure~\ref{fig:SRjets_PNNscores} for the $\tau_{\text{had}}$+jets SR and in Figure~\ref{fig:SRlep_PNNscores} for the $\tau_{\text{had}}$+lepton SRs.

\begin{table}[tp!]
\begin{center}
\caption{List of kinematic variables used as input to the PNN in the
$\tau_{\text{had}}$+jets and $\tau_{\text{had}}$+lepton channels.
Here, $\ell$ refers to the selected lepton (electron or muon),
while jet-1, jet-2 and jet-3 refer to the leading, second-leading and third-leading jet ordered in \pt.
The list of jets includes $b$-jets.
The $\Upsilon$ variable is sensitive to the polarization of
the $\tau$-lepton and is only defined for 1-prong \tauhadvis\
candidates. Hence, the PNN training is performed separately for
events with a selected 1- or 3-prong \tauhadvis\ candidate.}
\begin{minipage}{6cm}
\begin{tabular}{l|c|c}
PNN input variable & $\tau_{\text{had}}$+jets & $\tau_{\text{had}}$+lepton \\
\hline
$\pt^{\tau}$, $\eta^{\tau}$, $\phi^{\tau}$, $E^{\tau}$ & \checkmark & \checkmark \\
$\pt^{\ell}$, $\eta^{\ell}$, $\phi^{\ell}$, $E^{\ell}$ & & \checkmark \\
$\pt^{b\text{-jet}}$, $\eta^{b\text{-jet}}$, $\phi^{b\text{-jet}}$, $E^{b\text{-jet}}$ & \checkmark & \checkmark \\
$\pt^{\text{jet-1}}$, $\eta^{\text{jet-1}}$, $\phi^{\text{jet-1}}$, $E^{\text{jet-1}}$ & \checkmark & \checkmark \\
$\pt^{\text{jet-2}}$, $\eta^{\text{jet-2}}$, $\phi^{\text{jet-2}}$ & \checkmark & \\
$\pt^{\text{jet-3}}$, $\eta^{\text{jet-3}}$, $\phi^{\text{jet-3}}$ & \checkmark & \\
$\pt^{\text{jet-2}}$ & & \checkmark \\
$\met$, $\phi^{\met}$ & \checkmark & \checkmark \\
$\Upsilon$ (1-prong \tauhadvis only) & \checkmark & \checkmark \\
$m_{\text{truth}}^{H^+}$ & \checkmark & \checkmark \\
\end{tabular}
\end{minipage}
\label{Tab:pnn}
\end{center}
\end{table}


\subsection{Systematic uncertainties}

Detector-related systematic uncertainties arise from the simulation of the electron and muon triggers, from electron and muon isolation criteria, from the reconstruction and identification of leptons, $\tauhadvis$ objects and $b$-jets, from the energy/momentum scale and resolution of all detector objects, as well as from the reconstruction of the event $\met$.
To assess the impact of most detector-related systematic uncertainties on the four-momenta of the analysis objects, the selection criteria are re-applied after varying a particular parameter by its $\pm 1$ standard deviation value.
These are complemented by variations of the scale factors related to efficiencies of physics objects reconstruction and identification.
All detector-related systematic uncertainties are also propagated to
the analysis objects used in the $\met$ calculation, and an additional uncertainty on its soft term is considered~\cite{JETM-2020-03}.
The effect of each systematic uncertainty, with the exception of those that demonstrate less than $0.5\%$ deviation from the nominal value in all bins of the final discriminating variable as defined in Section~\ref{sec:result}, is parameterized and included in the definition of the likelihood function used to statistically infer the final results.

In the $\tau_{\text{had}}$+jets channel, additional uncertainties are attributed to the evaluation of the trigger efficiency scale factors.
These include a) the effect of varying the selection criteria used for measuring the trigger efficiency, such as changing the number of required jets and $b$ jets and applying different working points for the $\tauhadvis$ and electron identification algorithms; b) the impact of the fake $\tauhadvis$ background modeling, which is assessed by measuring the trigger scale factors after subtracting the data-driven fake $\tauhadvis$ estimate from the observed data and keeping only events with true $\tauhadvis$ in simulation; and c) statistical uncertainties in the values of the \emph{erf} function parameters derived from the fit.
The covariance matrix of the fitted parameters of the \emph{erf} function is diagonalized and then propagated to the final fit as uncorrelated uncertainties.

Uncertainties in the estimation of background from quark- and gluon-initiated jets misidentified $\tauhadvis$ objects arise from extraction of the FFs in the designated CRs as well as the determination of the combined FFs and application of those in the SRs.
Uncertainties in the FFs include statistical uncertainty in the FF value estimated in each \pt\ bin of the CRs, a 5\%\ uncertainty in the number of true \tauhad\ objects in MC simulation which satisfy the $\tauhadvis$ selection ($N_\mathrm{MCtrue}^{\tau\mathrm{-ID}}$) and a conservative 50\%\ uncertainty in the number of these objects passing the anti-$\tauhadvis$ selection ($N_\mathrm{MCtrue}^{\mathrm{anti-}\tau\mathrm{-ID}}$).
The former corresponds to the uncertainty in the \tauhadvis\ identification efficiency. There are no rigorous estimates of the latter, but the fractional contributions are small, as detailed in Section~\ref{subsec:bkg_modeling}.
The $\Upsilon$ variable transform derived from the FF $W$+jets CR is used as nominal while the one derived from the FF multi-jet CR is taken as a systematic uncertainty.
The uncertainties in determining the combined FFs include the statistical uncertainty in the template fit to the \tauhadvis\ jet width
as well as an uncertainty due to the \pt\ binning used for the template fit, which is estimated by varying the binning criteria.
Furthermore, the CRs used to measure FFs are mostly dominated by light-quark ($u, d, s$) and gluon initiated jets reconstructed as $\tauhadvis$, while in the SR with inverted $\tauhadvis$ identification criteria, around 30-40\%\ of the reconstructed $\tauhadvis$ candidates originate from heavy-flavor (HF) initiated jets.
The larger width of HF jets compared to light jets leads to a bias in the template fit and consequently in the composition of the jets misidentified as \tauhadvis\ objects in the SRs.
The bias is estimated in $t\bar{t}$ simulation by comparing the true yield of fake $\tauhadvis$ to the one estimated using FFs and jet width templates from light jets alone obtained from simulation.
The difference is used as an uncertainty on the fake $\tauhadvis$ yield from the $t\bar{t}$ background.
The largest single uncertainty on the fake $\tauhadvis$  estimation comes from the HF jet contribution in the SR.

Systematic uncertainties associated with reweighting of $t\bar{t}$ and $W$+jets MC samples include statistical uncertainties in the fitted values of the reweighting function parameters and the potential impact of signal contamination in CRs where reweighting factors are calculated.
To estimate the number of signal events, the $m_{H^+}=90~\GeV$ cross-section limit from Ref.~\cite{HIGG-2016-11} extrapolated\footnote{Scaling by the ratio of the expected limits at 80 and 90~\GeV.} to $m_{H^+}=80~\GeV$ is used.
The fractional contribution of the hypothetical signal amounts to 3.6\% (3.8\%) in the $\tau_{\text{had}}$+jets ($\tau_{\text{had}}$+lepton) $t\overline{t}$ CRs and 0.5\% in the $W$+jets CR.
An additional uncertainty associated with the extrapolation of the $t\bar{t}$ reweighting into the $\tau_{\text{had}}$+jets ($\tau_{\text{had}}$+lepton) SR is estimated by using the reweighting functions obtained for the $\tau_{\text{had}}$+lepton ($\tau_{\text{had}}$+jets) channel. In the case of the $W$+jets background, a conservative 50\% normalization uncertainty is considered.

Theoretical uncertainties on the main MC simulated backgrounds ($t\bar{t}$, single top quarks, $W$+jets) include contributions from renormalization and factorization scale variations, initial state radiation (ISR) and final state radiation (FSR), parton distribution functions, as well as parton shower and hadronization model. The latter is estimated using the difference between \Pythia8 (default) and \Herwig7~\cite{Bellm:2015jjp} predictions, which is then symmetrized. The prescriptions used for ISR/FSR and scale variations are generator dependent, but typically involve varying the resummation damping factor
up or down by a factor of two.
For each variation, a dedicated reweighting of the $t\bar{t}$ and $W$+jets MC samples is derived to avoid double counting of uncertainties.

Theoretical uncertainties on signal modeling include renormalization and factorization scale variations, PDF variations, as well as potential small acceptance loss due to the generator-level event selection used.
The PDF and scale variations are additionally rescaled to maintain the nominal cross-section.

The uncertainty in the integrated luminosity is 0.83\% \cite{DAPR-2021-01}, obtained using the LUCID-2 detector \cite{LUCID2} for the primary luminosity measurements, complemented by measurements using the inner detector and calorimeters. This uncertainty affects the normalization of all simulated samples.


%
%
%

%
\section{Results}
\label{sec:result}

%

%
\begin{table}[tp!]
\begin{center}
\caption{\label{tab:event_yields_SR_full} Pre-fit event yields for the backgrounds, hypothetical signal for two $\Hp$ mass hypotheses, and the observed number of data events in each of the three SRs.
The values shown for the signal assume $m_{H^+}$ = 170~\GeV\ and 1000~\GeV,
with a cross-section times branching fraction $\sigma(pp \to tb\Hp) \times \mathrm{\cal{B}}(\Hp \to \tau\nu)$ corresponding to 1~pb.
Combined statistical and systematic uncertainties are quoted.
}

\begin{tabular}{l|r|r|r}
& \multicolumn{3}{c}{Event yields and uncertainties} \\
Sample & \multicolumn{1}{c|}{$\tauhad$+jets} & \multicolumn{1}{c|}{$\tauhad+e$} & \multicolumn{1}{c}{$\tauhad+\mu$} \\
\hline
True $\tau_{\text{had}}$ & & & \\
\hspace*{5mm}$t\bar{t}$                  & $13400\phantom{.}\pm\phantom{.}2000$ & $39000\phantom{.}\pm\phantom{.}6000$ & $39000\phantom{.}\pm\phantom{.}5000$ \\
\hspace*{5mm}Single top quark            & $1600\phantom{.}\pm\phantom{.0}400$  & $2900 \phantom{.}\pm\phantom{.0}400$ & $3630 \phantom{.}\pm\phantom{.0}330$ \\
\hspace*{5mm}$Z \to \tau\tau$            & $220\phantom{.}\pm\phantom{.0}110$   & $900  \phantom{.}\pm\phantom{.0}500$ & $800  \phantom{.}\pm\phantom{.0}400$ \\
\hspace*{5mm}$W \to \tau\nu$             & $1770\phantom{.}\pm\phantom{.0}270$  & $2.4  \phantom{.}\pm\phantom{00}2.0$ & $0.07 \phantom{.}\pm\phantom{0}0.21$ \\
\hspace*{5mm}Diboson ($WW, WZ, ZZ$)      & $120\phantom{.}\pm\phantom{.00}60$   & $70   \phantom{.}\pm\phantom{.00}35$ & $80   \phantom{.}\pm\phantom{.00}40$ \\
Misidentified $\mbox{jet} \to \tauhadvis$ & $2430\phantom{.}\pm\phantom{.0}200$  & $8500 \phantom{.}\pm\phantom{.0}700$ & $8200 \phantom{.}\pm\phantom{.0}700$ \\
Misidentified $e,\,\mu \to \tauhadvis$    & $280\phantom{.}\pm\phantom{.0}170$   & $1000 \phantom{.}\pm\phantom{.0}500$ & $1000 \phantom{.}\pm\phantom{.0}500$ \\
\hline
All backgrounds                           & $19800\phantom{.}\pm\phantom{.}2100$ & $52000\phantom{.}\pm\phantom{.}6000$ & $53000\phantom{.}\pm\phantom{.}6000$ \\
\hline
$H^+$ $\phantom{0}$(170~\GeV), $\sigma \times \mathrm{\cal{B}}=1~\text{pb}$ & $980\phantom{.}\pm\phantom{.0}150$ & $580\phantom{.}\pm\phantom{.00}70$ & $690\phantom{.}\pm\phantom{.00}80$ \\
$H^+$ (1000~\GeV), $\sigma \times \mathrm{\cal{B}}=1~\text{pb}$              & $13000\phantom{.}\pm\phantom{.0}800$ & $940\phantom{.}\pm\phantom{.00}60$ & $1040\phantom{.}\pm\phantom{.00}70$ \\
\hline
Data         & $19650\phantom{.}\pm\phantom{.0}140$ & $51500\phantom{.}\pm\phantom{.0}230$ & $52730\phantom{.}\pm\phantom{.0}230$ \\
\end{tabular}
\end{center}
\end{table}

The expected pre-fit number of events for all SM processes and the measured event yields
in all three SRs are shown in Table~\ref{tab:event_yields_SR_full}.
The contributions from hypothetical
$H^+$ bosons are also listed, assuming a $m_{H^+}$ hypothesis of 170~\GeV\ or 1000~\GeV, and with
$\sigma(pp \to tbH^+) \times \mathrm{\cal{B}}(H^+ \to \tau\nu)$ set to
1~pb.
The signal selection acceptance times efficiency for a $m_{H^+}$ hypothesis of 170~\GeV, as
evaluated in an inclusive sample of simulated events where all possible $\tau$-lepton and top-quark decays are considered, is 0.78\%, 0.43\% and 0.51\% in the $\tau_{\text{had}}$+jets, $\tau_{\text{had}}$+electron and
$\tau_{\text{had}}$+muon SR, respectively. These become 9.35\%, 0.68\% and 0.74\%
for $m_{H^+} = $ 1000~\GeV. The large increase in the $\tau_{\text{had}}$+jets SR comes from the \MET\ requirement, which becomes almost fully efficient for large $m_{H^+}$.
The event yields measured in data are consistent with the SM background expectations in the three SRs.

The statistical interpretation is based on a simultaneous fit of the
parameter of interest, i.e.\ $\mu \equiv \sigma(pp \to tbH^+) \times
\mathrm{\cal{B}}(H^+ \to \tau\nu)$, and the nuisance parameters $\mathbf{\theta}$ that encode
statistical and systematic uncertainties, by means of a negative log-likelihood minimization.
The combined binned likelihood function ${\cal L}(\mu,\theta)$ for the PNN score distributions
is constructed as a product of Poisson probability terms over all bins in the three SRs ($\tau_{\text{had}}$+jets, $\tau_{\text{had}}$+electron and $\tau_{\text{had}}$+muon).
The binning of the discriminating variable is optimized to maximize the sensitivity of the analysis prior to looking at the data in the SRs.
The test statistic $\tilde{q}_{\mu}$~\cite{Cowan:2010js} used to test the compatibility of the
data with the background-only and signal+background hypotheses is computed from the profile likelihood ratio.
Upper limits on the signal production cross-section times branching fraction are derived using a binned likelihood fit with the CL$_{S}$ method~\cite{Read:2002hq}:
for a given signal hypothesis, values of the production cross-section times branching fraction
$\sigma(pp \to tbH^+) \times \mathrm{\cal{B}}(H^+ \to \tau\nu)$ yielding CL$_{S}$ < 0.05 are excluded
at 95\% CL. The asymptotic approximation~\cite{Cowan:2010js} is used throughout the statistical analysis.

\begin{figure}[tp!]
\centering

\subfloat[]{\includegraphics[width=0.45\linewidth]{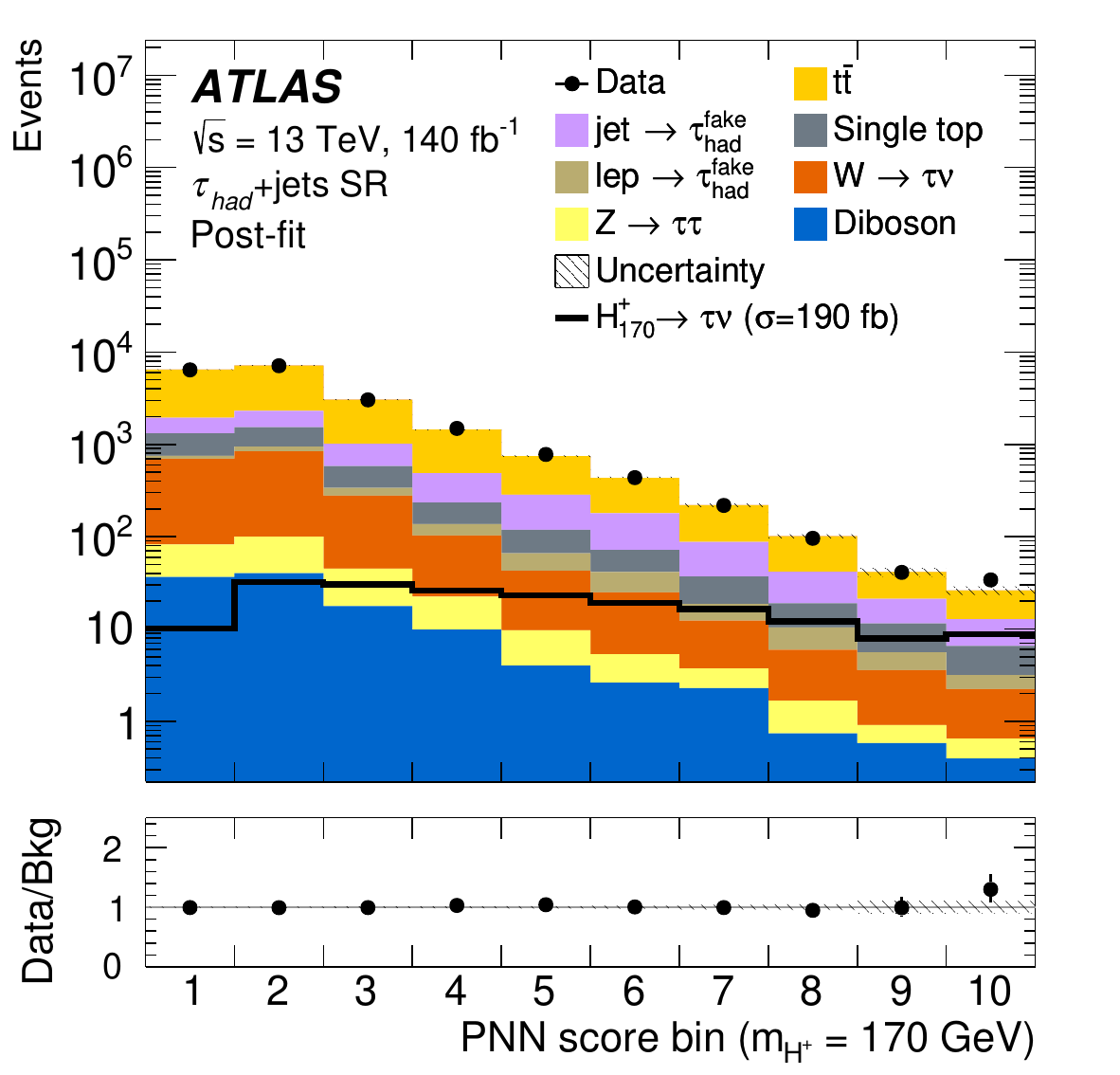}}
\subfloat[]{\includegraphics[width=0.45\linewidth]{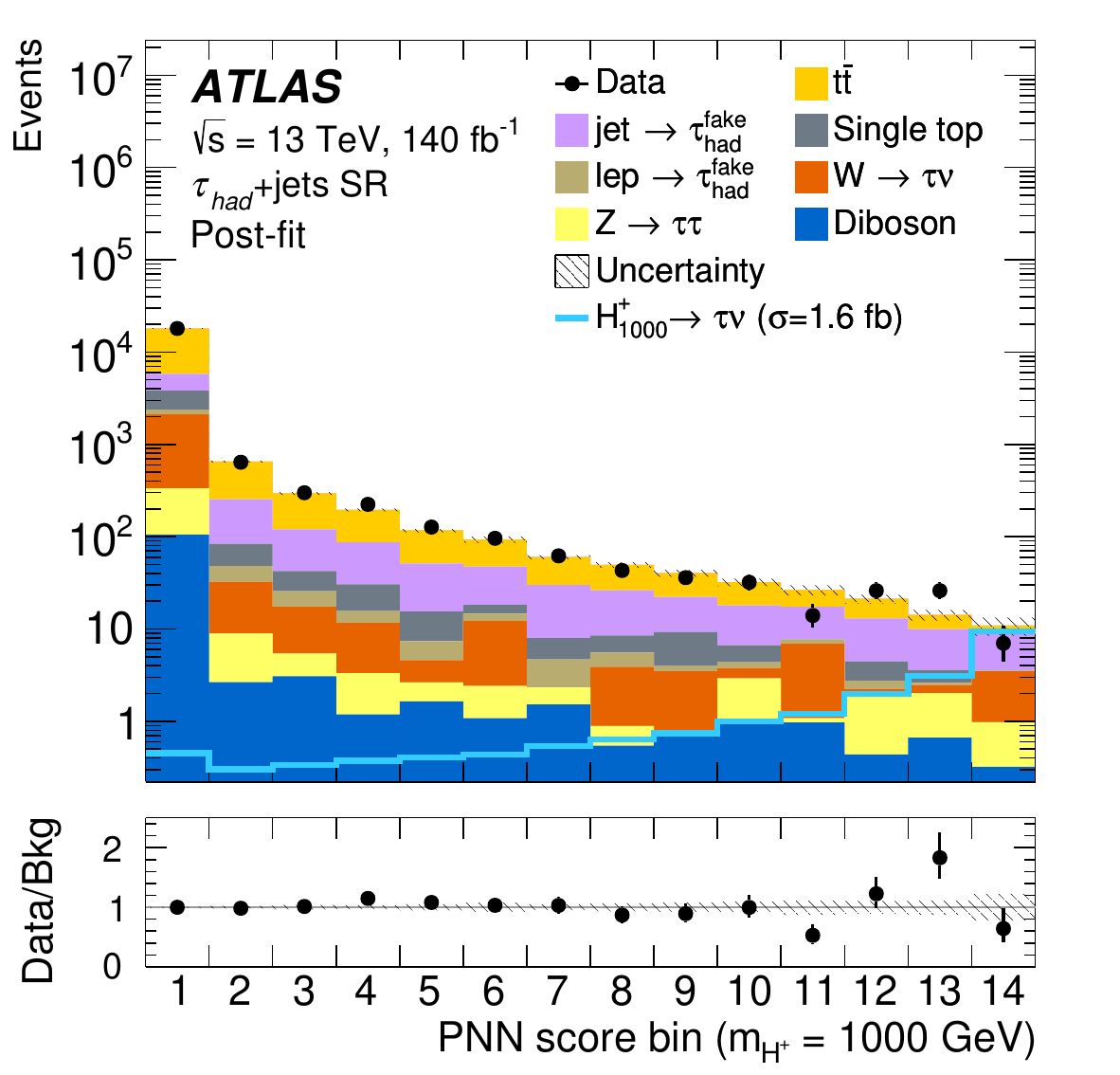}}

\caption{\label{fig:SRjets_PNNscores_body} PNN score distributions after the fit to data (``Post-fit'') under the background-only hypothesis in the
SR of the $\tau_{\text{had}}$+jets channel for the (a) 170~\GeV\ and (b) 1000~\GeV\ $m_{\Hp}$ hypotheses.
The signal distribution of the 170~\GeV\ (1000~\GeV) $\Hp$ mass hypothesis is shown by the solid black (teal) line, where the signal distribution is scaled to its expected 95\% CL limit. The lower panel of each plot shows the ratio of data to the SM background prediction.
The uncertainty bands include all statistical and systematic uncertainties.
Bin boundaries are optimized for the best sensitivity of the final fit (see Section~\ref{sec:result}) but are displayed as equidistant for better readability.
}
\end{figure}

\begin{figure}[t]
\centering
\subfloat[]{\includegraphics[width=0.45\linewidth]{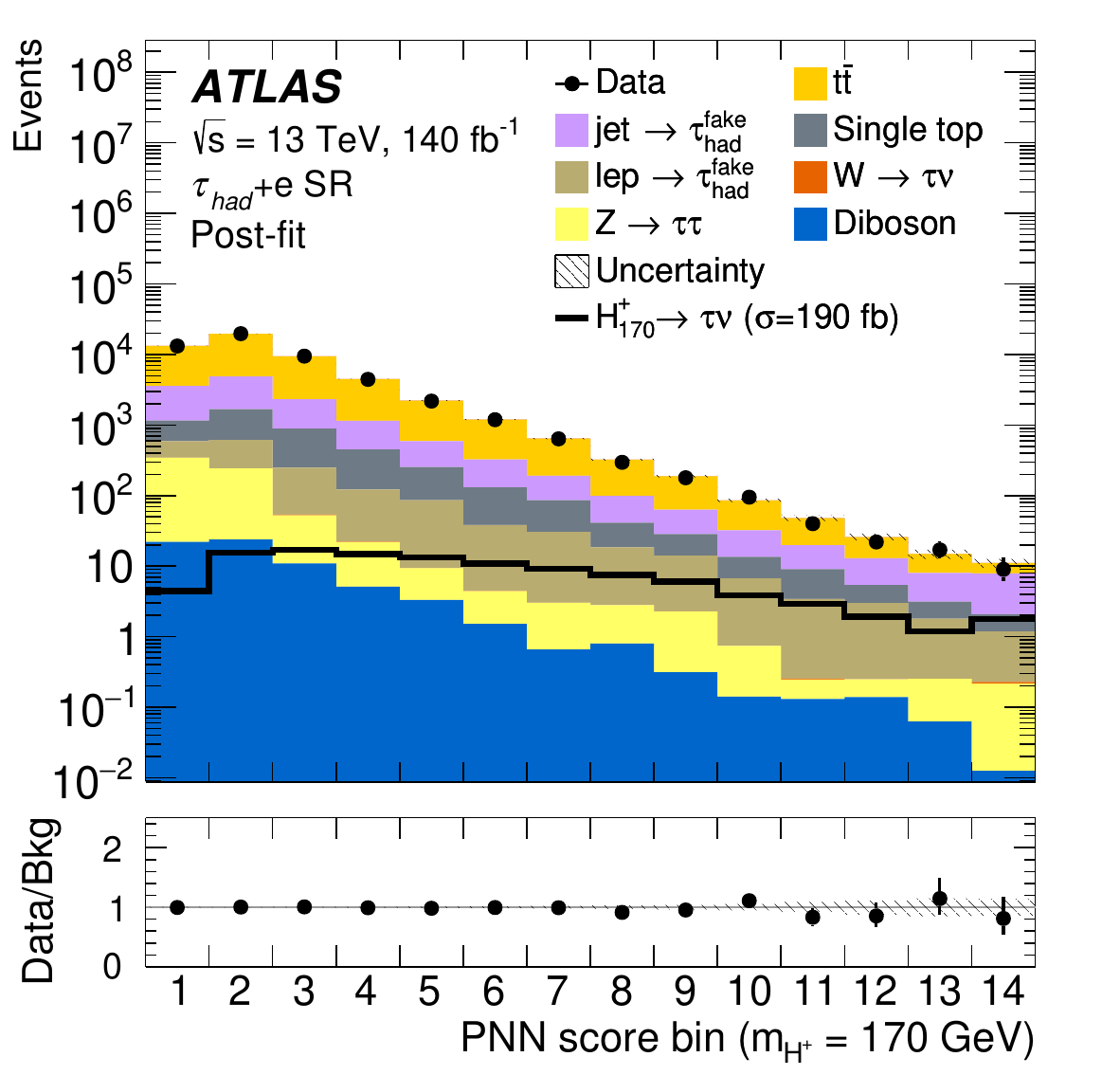}}
\subfloat[]{\includegraphics[width=0.45\linewidth]{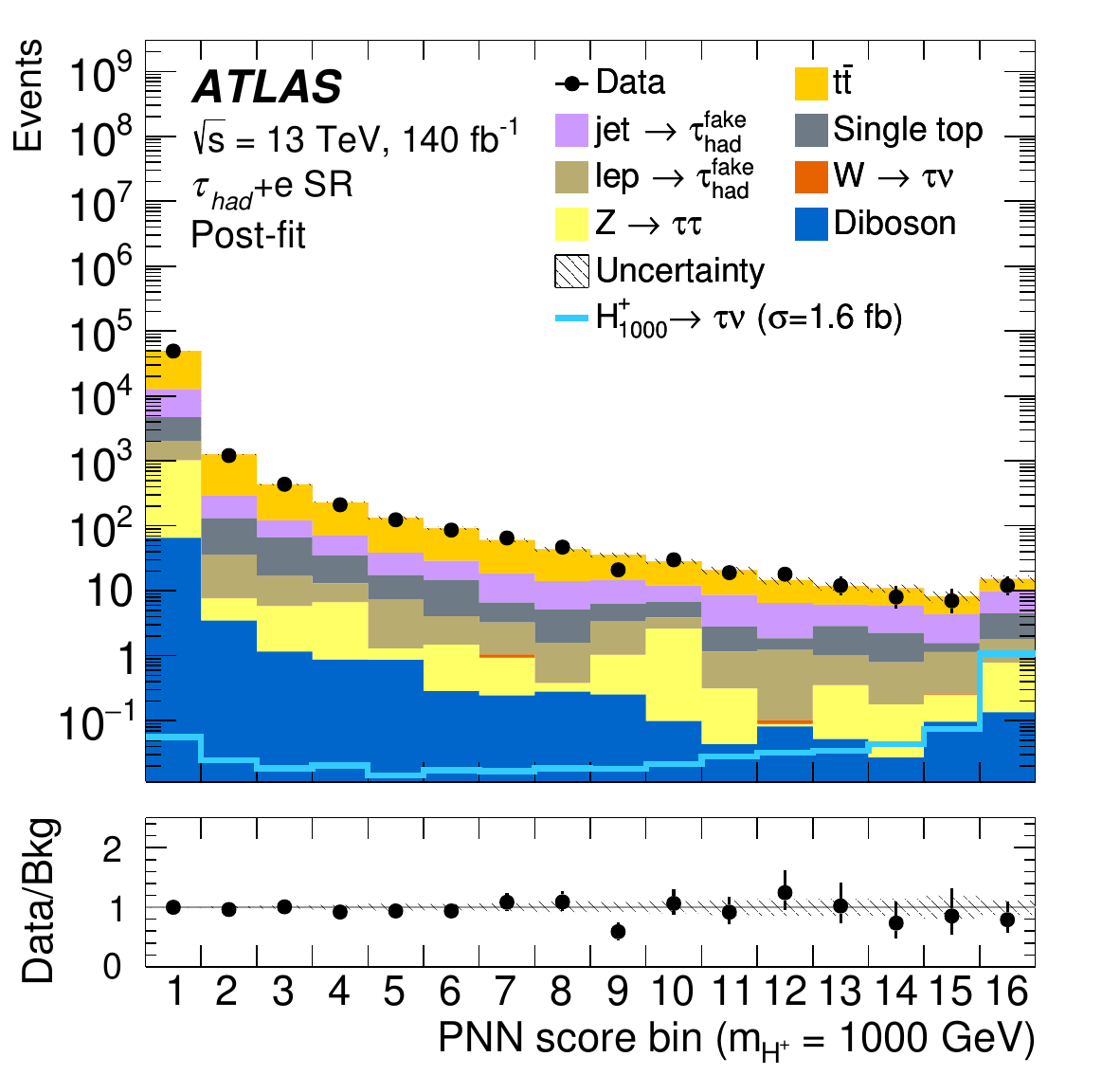}}   \\
\subfloat[]{\includegraphics[width=0.45\linewidth]{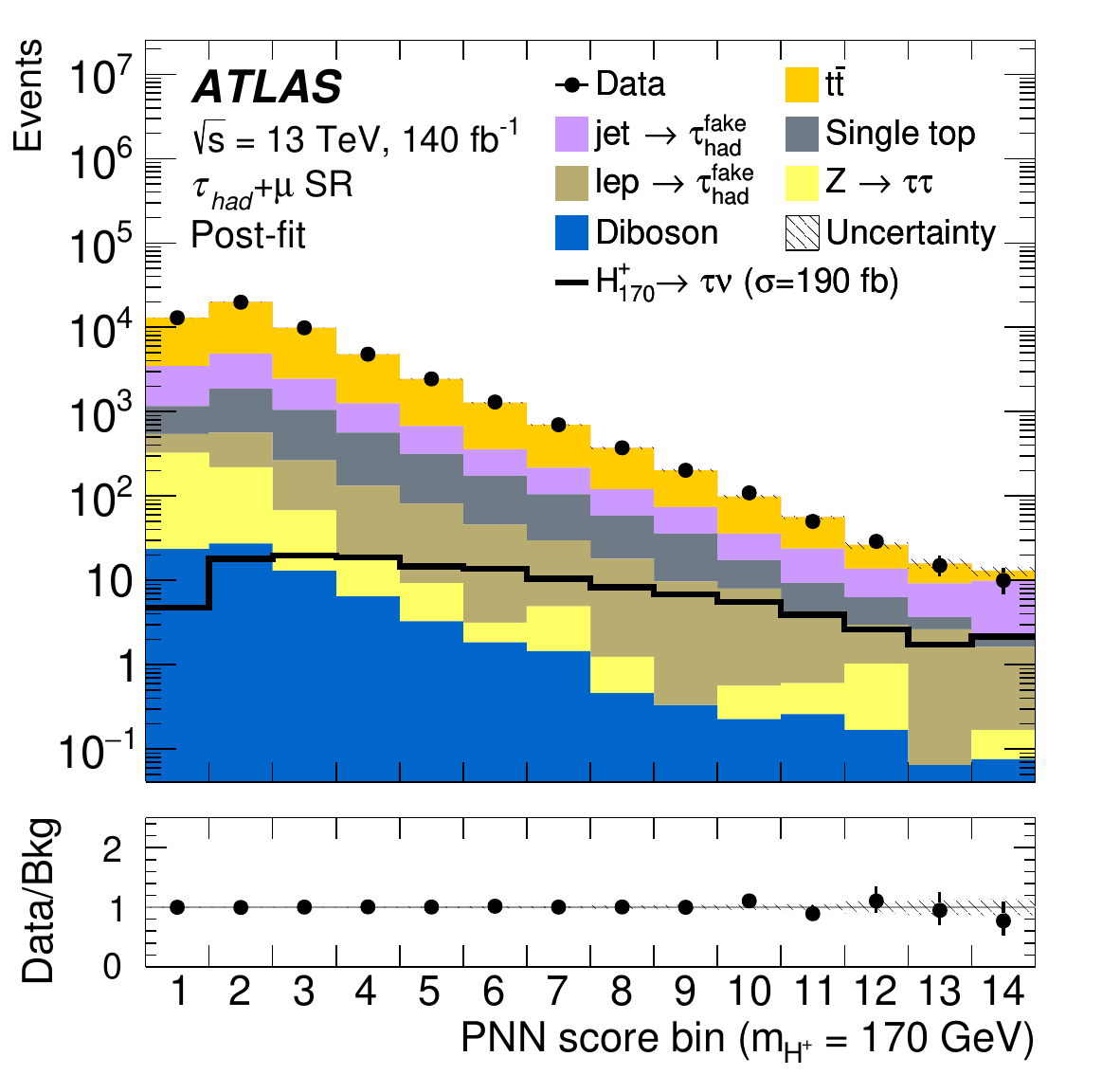}}
\subfloat[]{\includegraphics[width=0.45\linewidth]{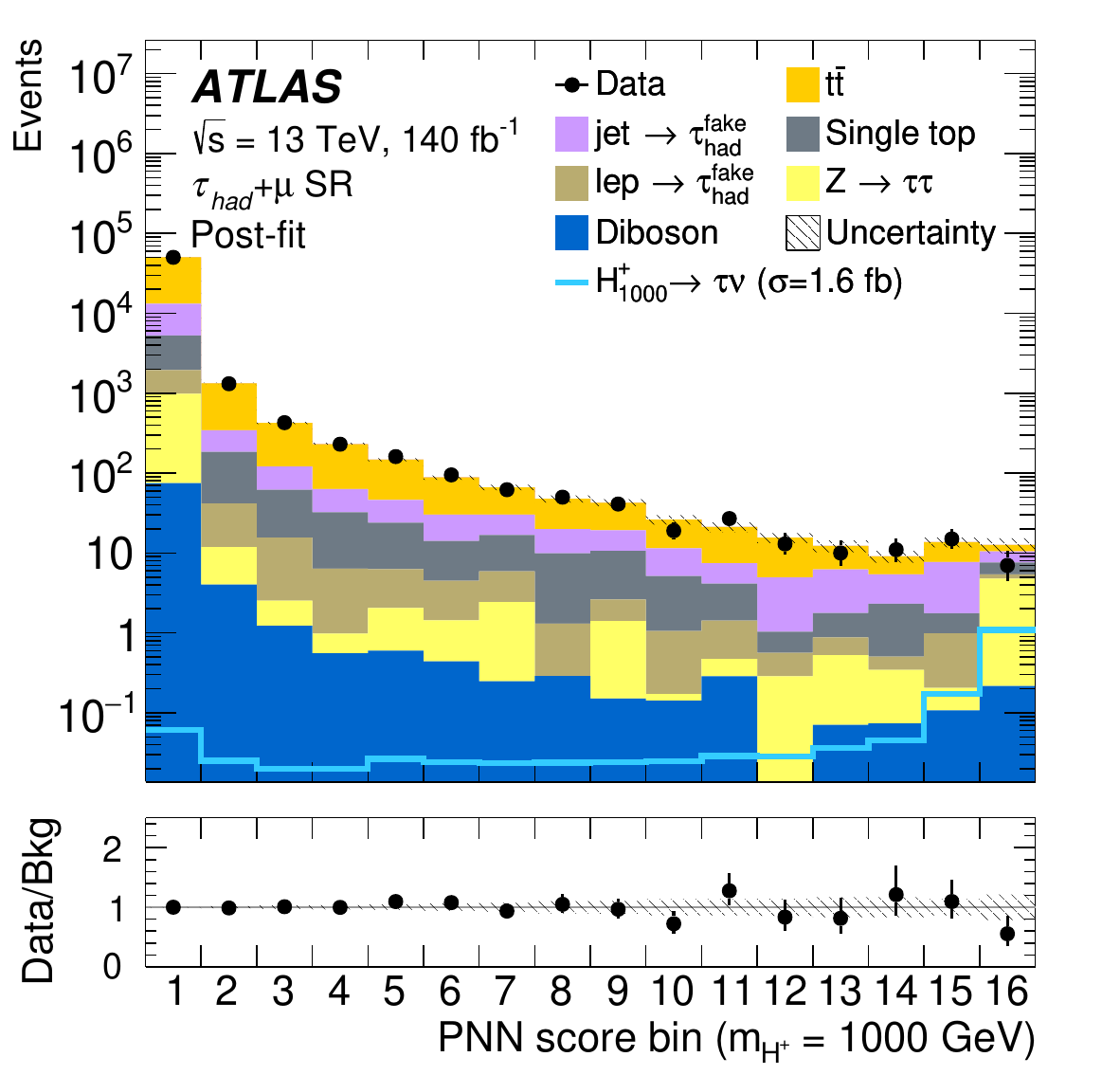}}

\caption{\label{fig:SRlep_PNNscores_body} PNN score distributions after the fit to data (``Post-fit'') under the background-only hypothesis in the SR of the $\tau_{\text{had}}$+electron sub-channel for the (a) 170~\GeV\ and (b) 1000~\GeV\ $m_{\Hp}$ hypotheses and $\tau_{\text{had}}$+muon sub-channel for the (c) 170~\GeV\ and (d) 1000~\GeV\ $m_{\Hp}$ hypotheses.
The signal distribution of the 170~\GeV\ (1000~\GeV) $\Hp$ mass hypothesis is shown by the solid black (teal) line, where the signal distribution is scaled to its expected 95\% CL limit. The lower panel of each plot shows the ratio of data to the SM background prediction.
The uncertainty bands include all statistical and systematic uncertainties.
Bin boundaries are optimized for the best sensitivity of the final fit (see Section~\ref{sec:result}) but are displayed as equidistant for better readability.
}
\end{figure}

Figures~\ref{fig:SRjets_PNNscores_body} and~\ref{fig:SRlep_PNNscores_body} show the PNN score distributions corresponding to two example $m_{H^+}$ hypotheses for the SRs of the $\tau_{\text{had}}$+jets and $\tau_{\text{had}}$+lepton channels, respectively.
All plots are obtained after the statistical fitting procedure with the background-only hypothesis.


\begin{figure}[t!]
\centering
\subfloat[]{\includegraphics[width=0.45\linewidth]{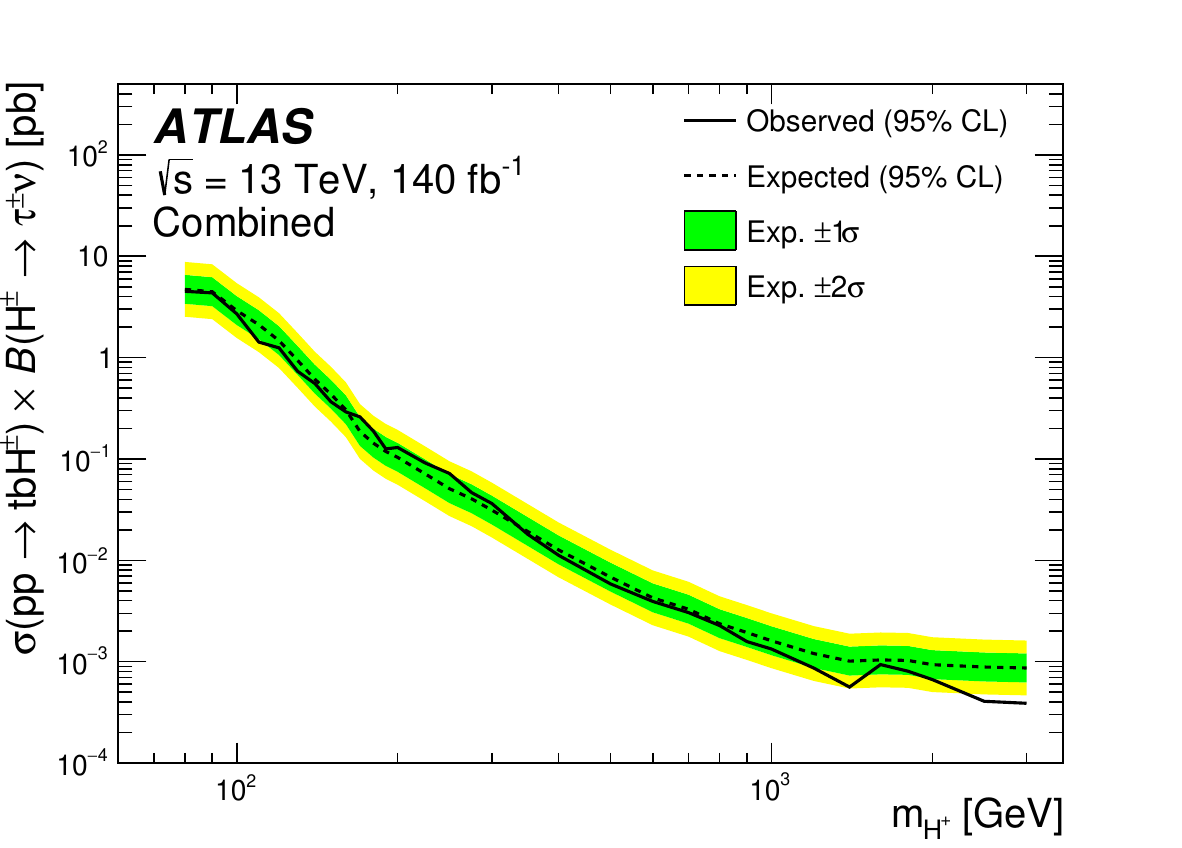}}
\subfloat[]{\includegraphics[width=0.45\linewidth]{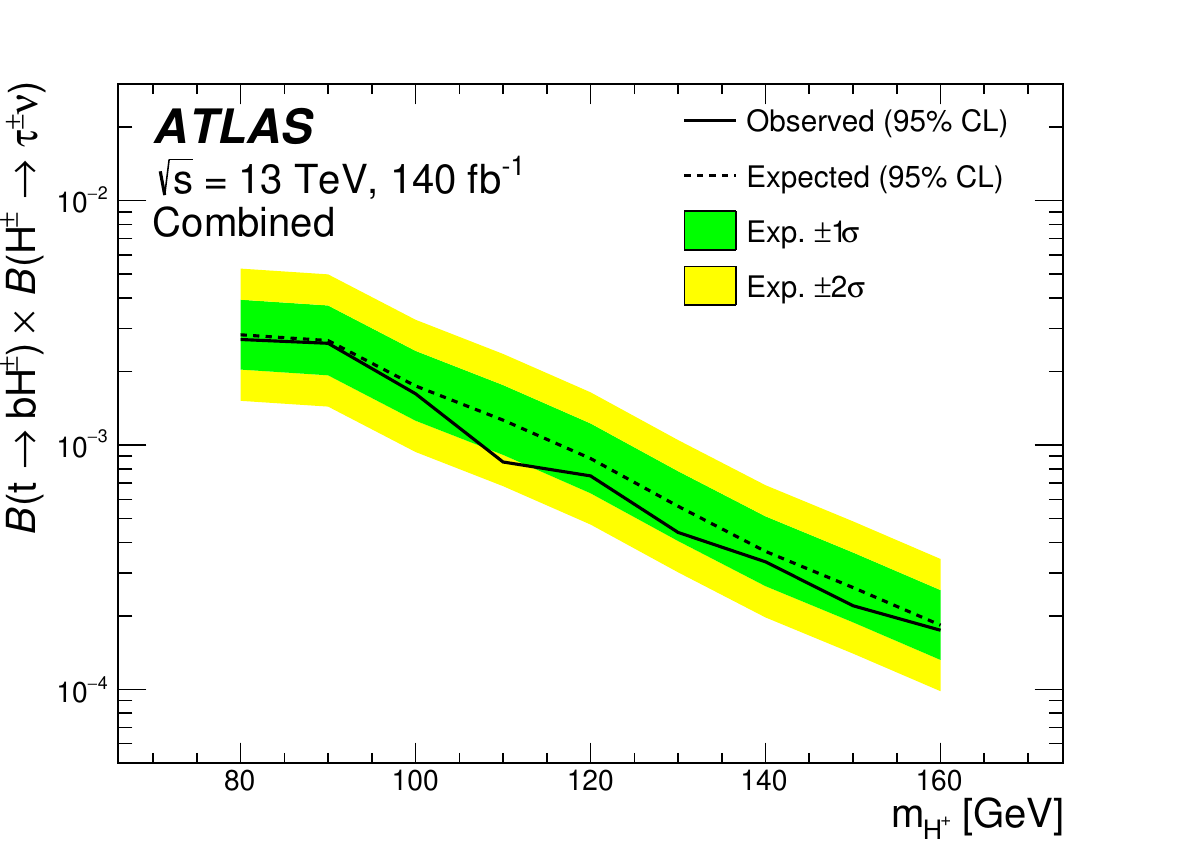}}
\caption{
Observed and expected 95\% CL exclusion limits on (a)~$\sigma(pp\to tbH^+)\times \mathrm{\cal{B}}(H^+ \to \tau \nu)$
and (b)~$\mathrm{\cal{B}}(t\to bH^+)\times \mathrm{\cal{B}}(H^+ \to \tau \nu)$ as a function of $m_{\Hp}$, from a combined fit in the \tauhadvis+jets and \tauhadvis+lepton channels.
The surrounding shaded bands correspond to the 1$\sigma$ and 2$\sigma$ confidence intervals around the expected limit.
}
\label{fig:main:limits}
\end{figure}

The data are found to be consistent under the background-only hypothesis.
Exclusion upper limits are set at 95\% CL on
$\sigma(pp \to tbH^+) \times \mathrm{\cal{B}}(H^+ \to \tau\nu)$ for the full mass range investigated,
as well as on $\mathrm{\cal{B}}(t\to bH^+)\times \mathrm{\cal{B}}(H^+ \to \tau \nu)$ in the low \Hp\ mass range.
Figure~\ref{fig:main:limits} shows the expected and observed combined exclusion limits
as a function of $m_{\Hp}$.
The observed limits range from 4.5~pb to 0.4~fb.
At low mass, the sensitivity of the analysis is driven by the $\tau_{\text{had}}$+lepton
channel, while at high mass the $\tau_{\text{had}}$+jets channel dominates.
For the mass range between 80 and 160~\GeV, the limits
on $\sigma(pp \to tb\Hp) \times \mathrm{\cal{B}}(\Hp \to \tau\nu)$ translate into observed limits between
0.27\% and 0.02\% on $\mathrm{\cal{B}}(t\to bH^+) \times \mathrm{\cal{B}}(H^+ \to \tau\nu)$
assuming the SM \ttbar\ production cross-section.

\begin{figure}[tb]
\centering
\subfloat[]{\includegraphics[width=0.45\linewidth]{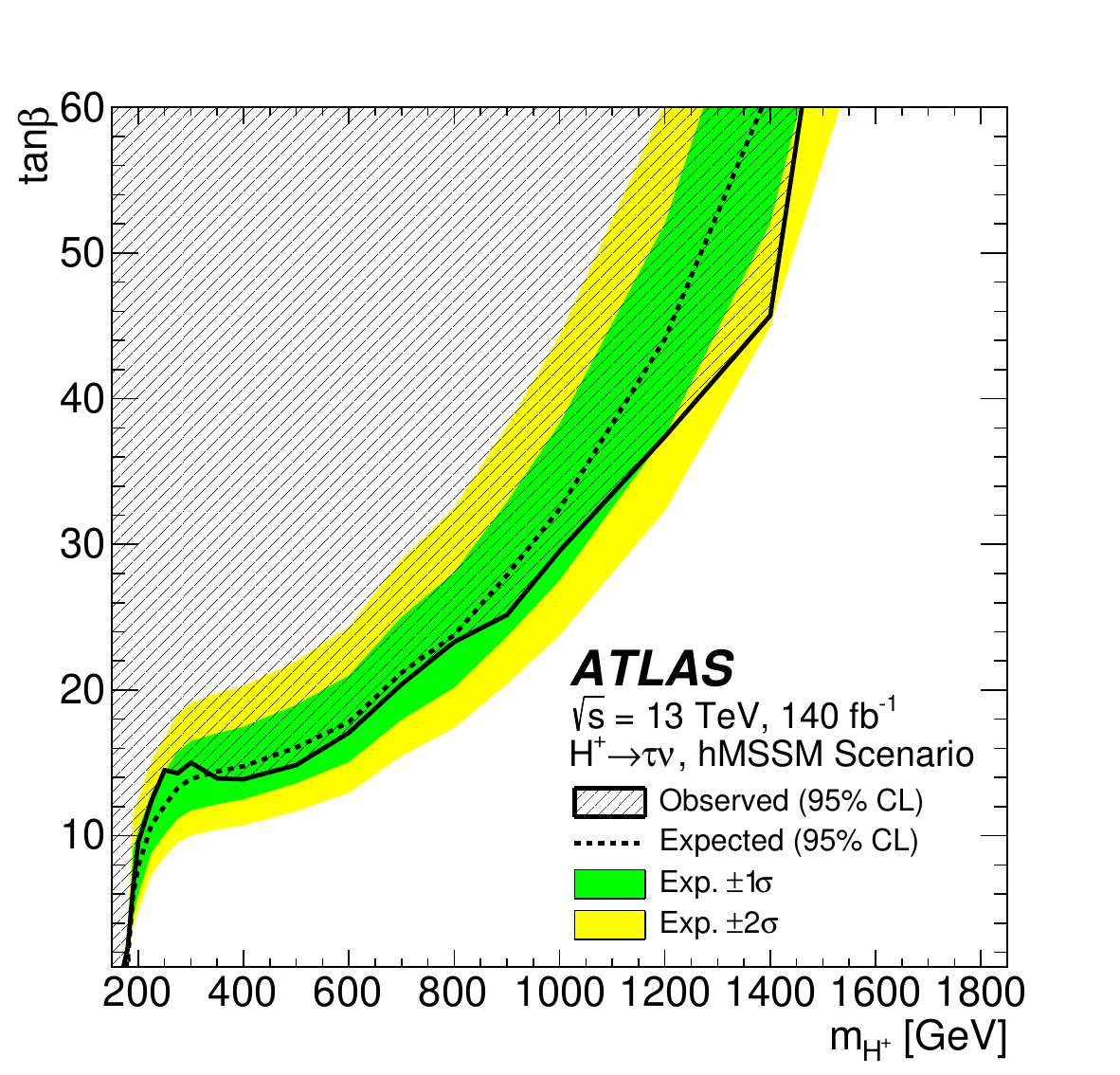}}
\subfloat[]{\includegraphics[width=0.45\linewidth]{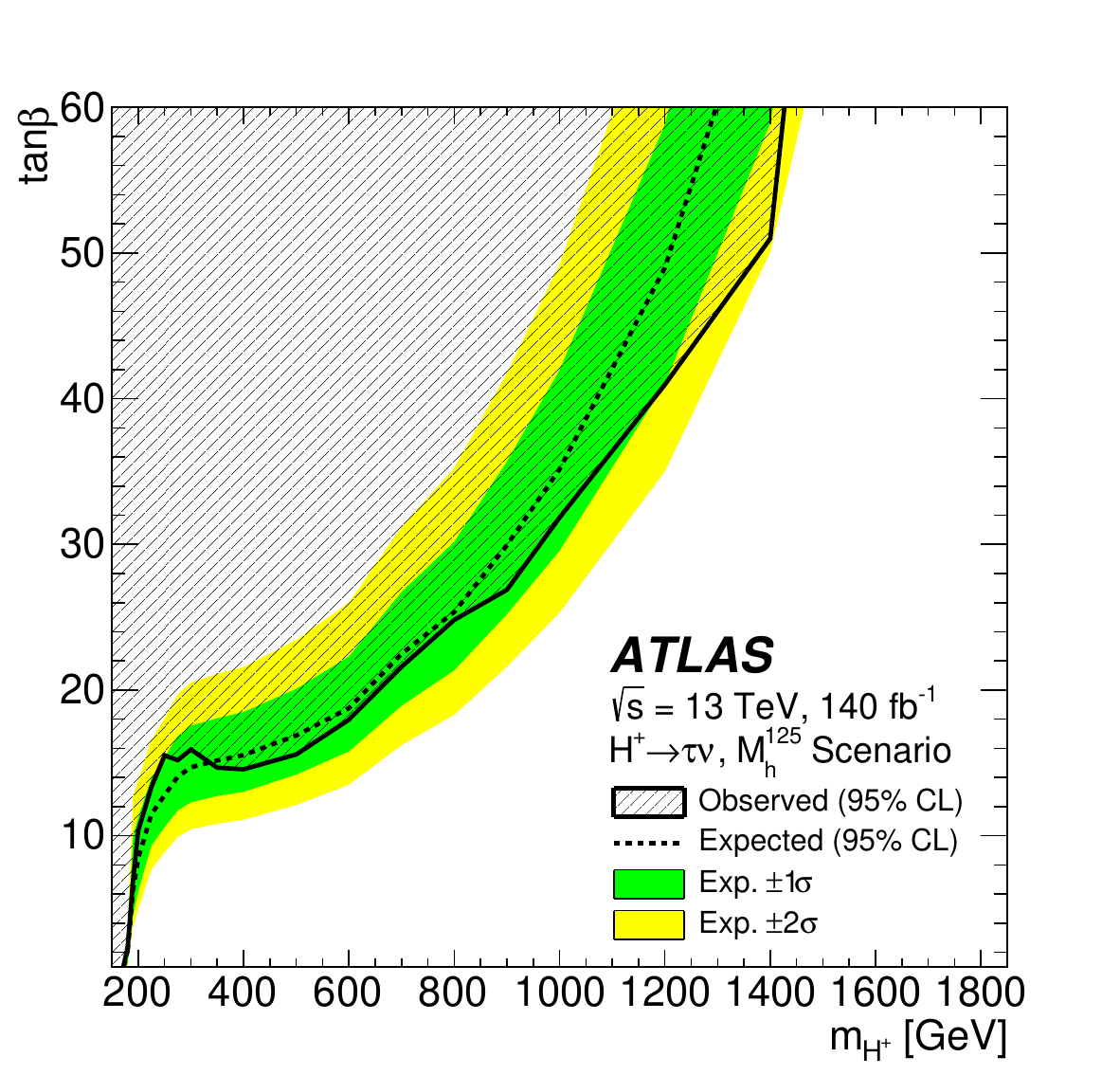}}
\caption{Observed and expected 95\% CL exclusion limits on $\tan\beta$ as a function of $m_{\Hp}$,
shown in the context of (a)~the hMSSM  and the (b)~the $M_h^{125}$ scenario of the MSSM, for $m_{\Hp}>150~\GeV$ and $(1 \leq \tan\beta \leq 60)$.
The surrounding shaded bands correspond to the 1$\sigma$ and 2$\sigma$ confidence intervals around the expected limit.
}
\label{fig:limits-interpreatation}
\end{figure}

The impact from the various sources of systematic uncertainty is estimated by taking the difference in quadrature of the observed 1 sigma uncertainty ($\sigma(\hat{\mu})$) on the best fit $\mu$ value ($\hat{\mu}$) obtained when performing an unconditional signal-plus-background fit to observed data while considering all systematic uncertainties,
to the $\sigma(\hat{\mu})$ obtained when performing a conditional fit with a certain set of systematic uncertainties fixed to their best fit values from the unconditional fit.
The results for example $m_{\Hp}$ mass hypotheses 170~\GeV\ and 1000~\GeV\ are summarized in Table~\ref{tab:impact_sys}.

Figure~\ref{fig:limits-interpreatation}
shows the 95\% CL exclusion limits on $\tan\beta$ as a function of $m_{\Hp}$ in the context of the hMSSM~\cite{Heinemeyer:2013tqa} and
the $M_h^{125}$ scenario of the MSSM~\cite{Heinemeyer:2019}.
The boundaries chosen correspond to regions in which theoretical predictions are available and in which this analysis has coverage.
Exclusion limits for $m_{\Hp}\leq140~\GeV$ are not shown, as the hMSSM scenario is not valid in this range~\cite{Djouadi:2015jea}.
At $\tan\beta = 60$,
above which no reliable theoretical calculations exist, $\Hp$ bosons
with masses up to 1400~\GeV\ are excluded, significantly improving on the limits
reported in Ref.~\cite{HIGG-2016-11}.
The hypothetical contribution from the $\Hp$ decaying to $tb$ with the subsequent tauonic top decay ($t\rightarrow b\tau^+\nu_{\tau}$) contributing to the low $\tan \beta$ region is small due to a low branching fraction of the semi-leptonic top-quark decay and has a lower reconstruction efficiency due to softer kinematics of the  $\tau\nu$ system.
Consequently, it is not considered in the derived exclusion, yielding a conservative result.

\begin{table}[tp!]
\begin{center}
\caption{\label{tab:impact_sys} %
Impact from various sources of systematic uncertainty on $\sigma(\hat{\mu})$ (the observed $1\sigma$ uncertainty on $\hat{\mu}$) from the combined fit for two $H^+$ mass
hypotheses: 170~\GeV\ and 1000~\GeV. The impact is obtained by taking the difference in quadrature of the uncertainty on $\hat{\mu}$ obtained when performing an unconditional signal+background fit to observed data ($\sigma(\hat{\mu}_{ALL})$), to the uncertainty on $\hat{\mu}$ obtained when performing a conditional fit with a certain set of systematic uncertainties fixed to their best fit values from the unconditional fit $\sigma(\hat{\mu}_{\mathrm{group}})$. The impact is defined as, Impact $= \frac{\sqrt{|\sigma(\hat{\mu}_{\mathrm{ALL}})|^2 - \sigma(\hat{\mu}_{\mathrm{group}})^2}\times100}{\sigma(\hat{\mu}_{\mathrm{ALL}})}$.
The row ``Total'' is obtained by computing the normalized difference in quadrature of the observed uncertainty from the unconditional fit, to a conditional fit where only statistical uncertainties are considered.
In the absence of correlations and assuming Gaussian uncertainties, the row ``Total'' would be obtained by summing in quadrature the individual contributions of the systematic uncertainties.\\}
\begin{tabular}{l|c|c}
Source of systematic & \multicolumn{2}{c}{Impact on $\sigma(\hat{\mu})$ [\%]} \\
uncertainty & $m_{H^{+}} = 170~\GeV$ & $m_{H^{+}} = 1000~\GeV$ \\
\hline
Experimental & & \\
\hspace*{5mm} Luminosity  &  $\phantom{>}$$\phantom{}1.0$   & $\phantom{>}$$\phantom{}1.0$ \\
\hspace*{5mm} Trigger     &  $\phantom{>}$$\phantom{}12$   & $\phantom{>}$$\phantom{}4.1$ \\
\hspace*{5mm} \tauhadvis\ &  $\phantom{>}$$\phantom{}12$   & $\phantom{>}$$\phantom{}8.3$ \\
\hspace*{5mm} Jet         &  $\phantom{>}$$\phantom{}27$   & $\phantom{>}$$\phantom{}17$ \\
\hspace*{5mm} Electron    &  $\phantom{>}$$\phantom{}1.9$   & $\phantom{>}$$\phantom{}4.0$ \\
\hspace*{5mm} Muon        &  $\phantom{>}$$\phantom{}9.3$   & $\phantom{>}$$\phantom{}2.2$ \\
\hspace*{5mm} $\met$      &  $\phantom{>}$$\phantom{}35$   & $\phantom{>}$$\phantom{}3.3$ \\
\hline
Fake-factor method         &  $\phantom{>}$$\phantom{}21$   & $\phantom{>}$$\phantom{}16$ \\
\hline
Signal and background models & & \\
\hspace*{5mm} $t\bar{t}$ modeling        &  $\phantom{>}$$\phantom{}20$      &  $\phantom{>}$$\phantom{}14$ \\
\hspace*{5mm} Single-top-quark modeling        &  $\phantom{>}$$\phantom{}46$      & $\phantom{>}$$\phantom{}2.6$ \\
\hspace*{5mm} $W/Z$+jets modeling        &  $\phantom{>}$$\phantom{}13$      &  $\phantom{>}$$\phantom{}51$ \\
\hspace*{5mm} Cross-sections $(W/Z/VV/t)$ &  $\phantom{>}$$\phantom{}2.3$      &  $\phantom{>}$$\phantom{}15$ \\
\hspace*{5mm} $H^{+}$ signal modeling    &  $\phantom{>}$$\phantom{}8.5$      &  $\phantom{>}$$\phantom{}1.0$ \\
\hline
Total                                 &  $\phantom{>}84$      & $\phantom{>}70$ \\
\end{tabular}
\end{center}
\end{table}


%

%
%
\FloatBarrier

\section{Conclusions}
%

A search for $\Hp$ bosons produced either in top-quark decays or in association with top quarks,
subsequently decaying via $\Hp \to \tau\nu$  is performed in the $\Hp$ mass range 80--3000~\GeV.
Depending on whether the top quark produced together with the $\Hp$ decays hadronically or semi-leptonically,
the search targets $\tau$+jets or $\tau$+lepton final states, in both cases with a $\tau$-lepton decaying into a neutrino and hadrons.
The dataset used for this analysis is from $pp$ collisions at $\sqrt{s}=13$~\TeV,
collected with the ATLAS detector at the LHC, and corresponds to an integrated luminosity of  \LUMI.
The discrimination between signal and background is based on a neural network parameterized as a function of the \Hp\ mass.

The data are found to be in agreement with the Standard Model background expectation.
Upper limits at 95\% CL are set on the $\Hp$ production cross-section times branching fraction, $\sigma(pp\rightarrow tb\Hp) \times \mathrm{\cal{B}}(\Hp \to \tau\nu)$, ranging from 4.5~pb to 0.4~fb for $\Hp$ masses in the range 80--3000~\GeV.
In the mass range 80--160~\GeV, assuming the SM cross-section
for \ttbar\ production, this corresponds to upper limits between 0.27\% and 0.02\% for the product of branching fractions
$\mathrm{\cal{B}}(t\to b\Hp) \times \mathrm{\cal{B}}(\Hp \to \tau\nu)$.
The upper limits are interpreted in the hMSSM and, for the first time in the $\tau\nu$ final state, the $M_h^{125}$ scenarios of the MSSM.
This search contributes to the broad program of \Hp\ searches at the LHC and provides the most restrictive limits to date in this production and decay mode.

%
%
%
%
%


%

%
\section*{Acknowledgments}
%

%
%

%
%

We thank CERN for the very successful operation of the LHC and its injectors, as well as the support staff at
CERN and at our institutions worldwide without whom ATLAS could not be operated efficiently.

The crucial computing support from all WLCG partners is acknowledged gratefully, in particular from CERN, the ATLAS Tier-1 facilities at TRIUMF/SFU (Canada), NDGF (Denmark, Norway, Sweden), CC-IN2P3 (France), KIT/GridKA (Germany), INFN-CNAF (Italy), NL-T1 (Netherlands), PIC (Spain), RAL (UK) and BNL (USA), the Tier-2 facilities worldwide and large non-WLCG resource providers. Major contributors of computing resources are listed in Ref.~\cite{ATL-SOFT-PUB-2025-001}.

We gratefully acknowledge the support of ANPCyT, Argentina; YerPhI, Armenia; ARC, Australia; BMWFW and FWF, Austria; ANAS, Azerbaijan; CNPq and FAPESP, Brazil; NSERC, NRC and CFI, Canada; CERN; ANID, Chile; CAS, MOST and NSFC, China; Minciencias, Colombia; MEYS CR, Czech Republic; DNRF and DNSRC, Denmark; IN2P3-CNRS and CEA-DRF/IRFU, France; SRNSFG, Georgia; BMBF, HGF and MPG, Germany; GSRI, Greece; RGC and Hong Kong SAR, China; ICHEP and Academy of Sciences and Humanities, Israel; INFN, Italy; MEXT and JSPS, Japan; CNRST, Morocco; NWO, Netherlands; RCN, Norway; MNiSW, Poland; FCT, Portugal; MNE/IFA, Romania; MSTDI, Serbia; MSSR, Slovakia; ARIS and MVZI, Slovenia; DSI/NRF, South Africa; MICIU/AEI, Spain; SRC and Wallenberg Foundation, Sweden; SERI, SNSF and Cantons of Bern and Geneva, Switzerland; NSTC, Taipei; TENMAK, T\"urkiye; STFC/UKRI, United Kingdom; DOE and NSF, United States of America.

Individual groups and members have received support from BCKDF, CANARIE, CRC and DRAC, Canada; CERN-CZ, FORTE and PRIMUS, Czech Republic; COST, ERC, ERDF, Horizon 2020, ICSC-NextGenerationEU and Marie Sk{\l}odowska-Curie Actions, European Union; Investissements d'Avenir Labex, Investissements d'Avenir Idex and ANR, France; DFG and AvH Foundation, Germany; Herakleitos, Thales and Aristeia programmes co-financed by EU-ESF and the Greek NSRF, Greece; BSF-NSF and MINERVA, Israel; NCN and NAWA, Poland; La Caixa Banking Foundation, CERCA Programme Generalitat de Catalunya and PROMETEO and GenT Programmes Generalitat Valenciana, Spain; G\"{o}ran Gustafssons Stiftelse, Sweden; The Royal Society and Leverhulme Trust, United Kingdom.

In addition, individual members wish to acknowledge support from Armenia: Yerevan Physics Institute (FAPERJ); CERN: European Organization for Nuclear Research (CERN DOCT); Chile: Agencia Nacional de Investigaci\'on y Desarrollo (FONDECYT 1230812, FONDECYT 1230987, FONDECYT 1240864); China: Chinese Ministry of Science and Technology (MOST-2023YFA1605700, MOST-2023YFA1609300), National Natural Science Foundation of China (NSFC - 12175119, NSFC 12275265, NSFC-12075060); Czech Republic: Czech Science Foundation (GACR - 24-11373S), Ministry of Education Youth and Sports (FORTE CZ.02.01.01/00/22\_008/0004632), PRIMUS Research Programme (PRIMUS/21/SCI/017); EU: H2020 European Research Council (ERC - 101002463); European Union: European Research Council (ERC - 948254, ERC 101089007, ERC, BARD, 101116429), European Union, Future Artificial Intelligence Research (FAIR-NextGenerationEU PE00000013), Italian Center for High Performance Computing, Big Data and Quantum Computing (ICSC, NextGenerationEU); France: Agence Nationale de la Recherche (ANR-20-CE31-0013, ANR-21-CE31-0013, ANR-21-CE31-0022, ANR-22-EDIR-0002); Germany: Baden-Württemberg Stiftung (BW Stiftung-Postdoc Eliteprogramme), Deutsche Forschungsgemeinschaft (DFG - 469666862, DFG - CR 312/5-2); Italy: Istituto Nazionale di Fisica Nucleare (ICSC, NextGenerationEU), Ministero dell'Università e della Ricerca (PRIN - 20223N7F8K - PNRR M4.C2.1.1); Japan: Japan Society for the Promotion of Science (JSPS KAKENHI JP22H01227, JSPS KAKENHI JP22H04944, JSPS KAKENHI JP22KK0227, JSPS KAKENHI JP23KK0245); Netherlands: Netherlands Organisation for Scientific Research (NWO Veni 2020 - VI.Veni.202.179); Norway: Research Council of Norway (RCN-314472); Poland: Ministry of Science and Higher Education (IDUB AGH, POB8, D4 no 9722), Polish National Agency for Academic Exchange (PPN/PPO/2020/1/00002/U/00001), Polish National Science Centre (NCN 2021/42/E/ST2/00350, NCN OPUS 2023/51/B/ST2/02507, NCN OPUS nr 2022/47/B/ST2/03059, NCN UMO-2019/34/E/ST2/00393, NCN \& H2020 MSCA 945339, UMO-2020/37/B/ST2/01043, UMO-2021/40/C/ST2/00187, UMO-2022/47/O/ST2/00148, UMO-2023/49/B/ST2/04085, UMO-2023/51/B/ST2/00920); Slovenia: Slovenian Research Agency (ARIS grant J1-3010); Spain: Generalitat Valenciana (Artemisa, FEDER, IDIFEDER/2018/048), Ministry of Science and Innovation (MCIN \& NextGenEU PCI2022-135018-2, MICIN \& FEDER PID2021-125273NB, RYC2019-028510-I, RYC2020-030254-I, RYC2021-031273-I, RYC2022-038164-I); Sweden: Carl Trygger Foundation (Carl Trygger Foundation CTS 22:2312), Swedish Research Council (Swedish Research Council 2023-04654, VR 2018-00482, VR 2022-03845, VR 2022-04683, VR 2023-03403, VR grant 2021-03651), Knut and Alice Wallenberg Foundation (KAW 2018.0458, KAW 2019.0447, KAW 2022.0358); Switzerland: Swiss National Science Foundation (SNSF - PCEFP2\_194658); United Kingdom: Leverhulme Trust (Leverhulme Trust RPG-2020-004), Royal Society (NIF-R1-231091); United States of America: U.S. Department of Energy (ECA DE-AC02-76SF00515), Neubauer Family Foundation.

%
%


%
%
%
%
%
%
%

%
\clearpage
\printbibliography
\clearpage
 
\begin{flushleft}
\hypersetup{urlcolor=black}
{\Large The ATLAS Collaboration}

\bigskip

\AtlasOrcid[0000-0002-6665-4934]{G.~Aad}$^\textrm{\scriptsize 104}$,
\AtlasOrcid[0000-0001-7616-1554]{E.~Aakvaag}$^\textrm{\scriptsize 17}$,
\AtlasOrcid[0000-0002-5888-2734]{B.~Abbott}$^\textrm{\scriptsize 123}$,
\AtlasOrcid[0000-0002-0287-5869]{S.~Abdelhameed}$^\textrm{\scriptsize 119a}$,
\AtlasOrcid[0000-0002-1002-1652]{K.~Abeling}$^\textrm{\scriptsize 56}$,
\AtlasOrcid[0000-0001-5763-2760]{N.J.~Abicht}$^\textrm{\scriptsize 50}$,
\AtlasOrcid[0000-0002-8496-9294]{S.H.~Abidi}$^\textrm{\scriptsize 30}$,
\AtlasOrcid[0009-0003-6578-220X]{M.~Aboelela}$^\textrm{\scriptsize 46}$,
\AtlasOrcid[0000-0002-9987-2292]{A.~Aboulhorma}$^\textrm{\scriptsize 36e}$,
\AtlasOrcid[0000-0001-5329-6640]{H.~Abramowicz}$^\textrm{\scriptsize 155}$
% ...
\bigskip
\\

% ...
$^{17}$Department for Physics and Technology, University of Bergen, Bergen; Norway.\\
% ...
$^{30}$Physics Department, Brookhaven National Laboratory, Upton NY; United States of America.\\
% ...
$^{36}$$^{(a)}$Facult\'e des Sciences Ain Chock, Universit\'e Hassan II de Casablanca;$^{(b)}$Facult\'{e} des Sciences, Universit\'{e} Ibn-Tofail, K\'{e}nitra;$^{(c)}$Facult\'e des Sciences Semlalia, Universit\'e Cadi Ayyad, LPHEA-Marrakech;$^{(d)}$LPMR, Facult\'e des Sciences, Universit\'e Mohamed Premier, Oujda;$^{(e)}$Facult\'e des sciences, Universit\'e Mohammed V, Rabat;$^{(f)}$Institute of Applied Physics, Mohammed VI Polytechnic University, Ben Guerir; Morocco.\\
% ...
$^{46}$Physics Department, Southern Methodist University, Dallas TX; United States of America.\\
% ...
$^{50}$Fakult\"{a}t Physik , Technische Universit{\"a}t Dortmund, Dortmund; Germany.\\
% ...
$^{56}$II. Physikalisches Institut, Georg-August-Universit\"{a}t G\"ottingen, G\"ottingen; Germany.\\
% ...
$^{104}$CPPM, Aix-Marseille Universit\'e, CNRS/IN2P3, Marseille; France.\\
% ...
$^{119}$$^{(a)}$New York University Abu Dhabi, Abu Dhabi;$^{(b)}$United Arab Emirates University, Al Ain; United Arab Emirates.\\
% ...
$^{123}$Homer L. Dodge Department of Physics and Astronomy, University of Oklahoma, Norman OK; United States of America.\\
% ...
$^{155}$Raymond and Beverly Sackler School of Physics and Astronomy, Tel Aviv University, Tel Aviv; Israel.\\
% ...
$^{a}$ Also Affiliated with an institute covered by a cooperation agreement with CERN.\\
$^{b}$ Also at An-Najah National University, Nablus; Palestine.\\
$^{c}$ Also at Borough of Manhattan Community College, City University of New York, New York NY; United States of America.\\
$^{d}$ Also at Center for High Energy Physics, Peking University; China.\\
$^{e}$ Also at Center for Interdisciplinary Research and Innovation (CIRI-AUTH), Thessaloniki; Greece.\\
$^{f}$ Also at CERN, Geneva; Switzerland.\\
% ...

\end{flushleft}


%

%
%

%

%
%
%
%

\end{document}